\documentclass[pra,twocolumn,showpacs,floatfix,amsmath,amsfonts]{revtex4}
\usepackage{bm}
\usepackage{graphicx}

\newcommand{\bra}[1]{\langle #1 | \,}
\newcommand{\ket}[1]{\, | #1 \rangle}
\newcommand{\braket}[2]{\langle #1 | #2 \rangle}
\newcommand{\la}{\lambda}

\newcommand{\ga}{\gamma}

\newcommand{\de}{\delta}
\newcommand{\De}{\Delta}
\newcommand{\e}{\epsilon}
\newcommand{\ve}{\varepsilon}
\newcommand{\us}{\uparrow}
\newcommand{\ds}{\downarrow}
\newcommand{\ak}{a^{\dagger}}
\newcommand{\be}{\begin{equation}}
\newcommand{\ee}{\end{equation}}
\newcommand{\bea}{\begin{eqnarray}}
\newcommand{\eea}{\end{eqnarray}}

\begin{document}

\title{Electron wavepacket propagation and entanglement in a chain of 
coupled quantum dots}

\author{Georgios M. Nikolopoulos}
\altaffiliation[Present address: ]{Institut f\"ur Angewandte Physik, 
Hochschulstrasse 4a, 64289 Darmstadt, Germany}
\affiliation{Institute of Electronic Structure \& Laser, FORTH, P.O.
 Box 1527, Heraklion 71110, Crete, Greece}
\author{David Petrosyan} 
\affiliation{Institute of Electronic Structure \& Laser, FORTH, P.O.
 Box 1527, Heraklion 71110, Crete, Greece}
\author{P. Lambropoulos} 
\altaffiliation[Also at the ]{Department of Physics, University of Crete, 
Greece}
\affiliation{Institute of Electronic Structure \& Laser, FORTH, P.O.
 Box 1527, Heraklion 71110, Crete, Greece}

\date{\today}

\begin{abstract}
We study the coherent dynamics of one- and two-electron transport in a 
linear array of tunnel-coupled quantum dots. We find that this system 
exhibits a rich variety of coherent phenomena, ranging from electron 
wavepacket propagation and interference to two-particle bonding and 
entanglement. Our studies, apart from their relevance to the exploration 
of quantum dynamics and transport in periodic structures, are also aimed 
at possible applications in future quantum computation schemes.
\end{abstract}


\maketitle

\section{Introduction}

Quantum transport is by now an almost generic term encompassing a broad
class of phenomena in normally quite distinct areas of physics. 
Traditionally the term was found in solid state physics \cite{solst}, 
while through developments over the last several years, it is commonly 
found in the context of nonlinear dynamics pertaining to quantum chaos 
\cite{Qchaos}, in cold atoms in optical lattices \cite{mat_wave1,mat_wave2}, 
Rydberg atoms strongly  driven by microwave fields \cite{andreas}, or even 
laser-driven multilevel systems \cite{shore1,shore2,shore3}, the latter 
dating back twenty five years or so, although the term was never used in 
that context. Issues of coherent propagation of wavepackets, their interplay 
with disorder and/or fluctuations, localization, etc. are now discussed in 
many of these contexts and often interchangeably \cite{anderson}.

A new context, namely low-dimensional nanostructures and in particular
quantum dots (QD) \cite{QDots,ElTrQD} has emerged over the last ten or so 
years which appears to overlap with and in some sense unify several of these 
areas. Often referred to as artificial atoms \cite{KAT,ReMa}, they offer 
the unprecedented possibility to construct at will and explore situations 
ranging from practically single atom to a fully solid state many-body systems
\cite{WDFEFTK,VGNHMMF,KHvWHTF,SDS,KSDS,SKDS,GuPr,Gurvitz,WeNa,KChY,YJH}.
The nanofabrication possibilities of tailoring structures to desired
geometries and specifications, and controlling the number and mobility
of electrons confined within a region of space 
\cite{KAT,ReMa,EHGetc,VHWvBetc}, are some of the features that make these 
structures unique tools for the study of  a variety of preselected  set of 
phenomena, not to mention the potential for devices. 
Given the controllable quantum properties of the electrons in such
confined structures, the possibility of their application to schemes of
quantum computers \cite{QCI,RevSteane} has not escaped attention as attested 
by an ever growing plethora of papers on this general issue 
\cite{LDV,BLDV,EHGetc,VHWvBetc,zanros,LiAra,Tanamoto,BHHFKWSF}.

Motivated in part by the connection with quantum computing and in part by
the desire to explore the dynamics of wavepackets in periodic structures, 
we began with the study of such wavepackets in the simplest case of one
electron tunneling through an array of QDs.  As discussed in some detail
below, the problem turns out to have unexpected similarities with an old
problem, namely the molecular excitation through a ladder of vibrational
levels \cite{shore1,shore2,shore3}. Models developed in that context 
more than twenty five years ago found no application, simply because a 
vibrational structure is too complex to be amenable to simple models of, 
for example, equidistant levels or exact levels of a harmonic oscillator. 
On the other hand, in the present case, one can ~easily contemplate 
fabricating almost any desired level and tunneling rates arrangement. 
Thus, models that were of only academic interest then, now become of 
practical relevance. At the same time,  the equations governing the 
propagation in a chain of quantum dots are similar to those found in 
the 1-dimensional (1D) Ising model \cite{ising}.

Even a slight generalization, namely the presence of one more excess
electron in the chain presents an entirely different situation. Its
counterpart in multilevel systems does not exist, while it is now the 2D 
Ising model \cite{ising} that is related to the chain of QDs. In our case, 
however, it is the dynamical rather than the thermodynamic properties of the
system that we explore. With two electrons available for tunneling, an
additional set of phenomena such as Coulomb blockade, bonding (to be
described below), and correlation come into play. Surprisingly, we
find that preservation of coherence found in the one-electron case, under
a judiciously chosen set of parameters, carries through in the two-electron 
case. It is this finding that brings in the possible connection with schemes 
for the quantum computer. Such a possibility could be contemplated when the 
electron spin is brought into the picture, combined with coherent propagation 
within a chain of QDs ~and manipulation of spin-entanglement \cite{LDV,BLDV},
as discussed in Sec. \ref{sec:etngl}.

Even with the high degree of accuracy in nanofabrication technology, issues
of randomness and fluctuations \cite{QDots,phonons}, originating from ~more 
than one physical effect, are an inevitable part of the picture. Their 
presence, in addition to requiring attention with respect to their impact 
on coherent propagation and entanglement, represents a class of further 
phenomena related to the well-known localization induced by disorder, 
and related issues \cite{anderson}. We have therefore examined some of 
our basic results in the presence of randomness. Although not necessary 
for the main arguments of this paper, for the sake of completeness,
we have also explicitly included in our analysis the effect of measurement. 
This has necessitated the adoption of Monte Carlo techniques which are well
established in quantum optics  and provide a direct account of the effect
of a measurement on the system \cite{DCM,DZR,RevQJumps}.

We would like to state at the outset that, although we do realize that
each quantum dot in general has many electrons, it is nevertheless
possible to arrange the conditions so that only one electron per dot
tunnels, while the rest remain ``frozen''. It may not be routine, but 
on the basis of existing literature \cite{HDS,ElTrQD,ReMa,WDFEFTK}, 
it is technologically possible.  In fact many papers with proposals 
for quantum registers employing QDs are based on that very idea 
\cite{LDV,LiAra,VHWvBetc,Tanamoto}. We do also realize that assuming a 
chain of 10 or 20 dots, as we do in the sections to follow, may indeed 
stretch present day technological capabilities. We do so, however, in the
interest of exploring a number of unusual effects which if not quite 
feasible today will most likely be feasible tomorrow.  But when it comes 
to entanglement in connection with a quantum register, in the last section 
of the paper, even two dots fulfill our requirements.

We have structured this paper as follows: In Sec. \ref{sec:mform} we
present the mathematical formalism describing 1D chains of QDs, upon 
which we build the theory of coherent single-electron propagation in 
Sec. \ref{sec:1e} and two-electron propagation in Sec. \ref{sec:2e}.
In Sec. \ref{sec:etngl} we show a scheme for entangling two electrons
in the chain via their controlled collision. Our conclusions are 
summarized in Sec. \ref{sec:concl}.

\section{Mathematical formalism}
\label{sec:mform}
The system under consideration consists of a linear array of $N$ nearly 
identical QDs which are electrostatically defined in a two-dimensional 
electron gas by means of metallic gates on top of a semiconductor 
heterostructure, e.g., GaAs/AlGaAs \cite{QDots,KAT,ElTrQD}. We 
describe the evolution of the system using the second-quantized, 
extended Mott-Hubbard Hamiltonian \cite{SDS,GuPr,WeNa}
\bea
H &=& \sum_{j,\alpha} \ve_{j\alpha} \ak_{j\alpha} a_{j\alpha}
+\frac{1}{2} \sum_{j} U n_j (n_j -1) \nonumber \\ & &
+\sum_{i < j,\alpha} t_{ij,\alpha} (\ak_{i\alpha} a_{j\alpha}+{\rm H.c.})
+ \sum_{i < j} V_{ij} n_i n_j ,
\label{Ham}
\eea
the first two terms being responsible for the intradot effects and 
the last two terms describing the interdot interactions. Here
$\ak_{j\alpha}$ and $a_{j\alpha}$ are the creation and annihilation 
operators for an electron in state $\alpha$ with the single-particle energy 
$\ve_{j\alpha}$ and electronic orbital $\psi_{j\alpha} (\mathbf{r})$, 
$U = \frac{e^2}{8 \pi \e_r \e_0} \int d \mathbf{r} d \mathbf{r}^{\prime}
|\psi_{j\alpha} (\mathbf{r})|^2|\psi_{j\alpha^{\prime}} 
(\mathbf{r}^{\prime})|^2 /| \mathbf{r} - \mathbf{r}^{\prime}| \simeq e^2/C_g$,
with $C_g \simeq 8 \e_r \e_0 R$ being the self-capacitance for 2D disk-shaped 
QD ($\e_r \simeq 13$ for GaAs), is the on-site Coulomb repulsion, 
$n_j=\sum_{\alpha} \ak_{j\alpha} a_{j\alpha}$ is the total electron number 
operator of the $j$th dot, $t_{ij,\alpha} = \frac{\hbar^2}{2 m^{*}}  
\int d \mathbf{r} \psi_{i\alpha}^{*} (\mathbf{r}) 
\nabla^2 \psi_{j\alpha} (\mathbf{r})$, with $m^{*}$ being the electron 
effective mass ($m^{*} \simeq 0.067 m_e$ in GaAs), are the coherent tunnel 
matrix elements which are given by the overlap of the electronic orbitals 
of adjacent dots ($j=i + 1$) and thereby can be controlled by the external 
voltage applied to the gates defining the corresponding tunneling barriers 
between the dots, and $V_{ij}\simeq U (C/C_g)^{|i-j|}$, with $C \ll C_g$ 
being the interdot capacitance, describe the interdot 
electrostatic interaction. Note that in general, the index $\alpha$ refers to 
both orbital and spin state of an electron. In the Coulomb blockade and
tight-binding regime, when the on-site Coulomb repulsion and single-particle 
level-spacing $\De \ve \simeq \frac{\hbar^2 \pi}{m^{*} R^2}$ are much larger 
than the tunneling rates, $U > \De \ve \gg t_{ij,\alpha}$ \cite{cmnt}, only 
the equivalent states of the neighboring dots are tunnel-coupled to each 
other. Therefore throughout this paper we will consider a single doubly- 
(spin-) degenerate level per dot ($\alpha\in\{\us,\ds\}$), assuming further 
that the tunneling rates do not depend on the electron spin 
($t_{ij,\alpha} = t_{ij}$). In our subsequent discussion, we assume that 
the matrix elements $V_{ij} = V$ are non-vanishing for nearest neighbors only,
$j=i+1$. Interdot repulsion can be further suppressed in the presence 
of a nearby conducting backgate, where image charges of excess electrons 
completely screen the interdot Coulomb repulsion, in which case $V \approx 0$.

\begin{figure}[t]
\centerline{\includegraphics[width=8.5cm]{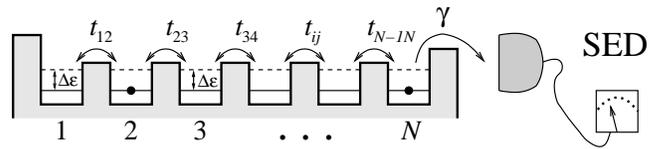}}
\caption{Schematic drawing of the chain of tunnel-coupled QDs, 
and the last dot is dissipatively coupled to an external 
single electron detector (SED).
\label{scheme}}
\end{figure}

We study situations where one or more preselected QDs are initially doped 
with single electrons (Fig. \ref{scheme}). The preparation of such initial 
conditions can be accomplished by first applying a large negative voltage 
to the gates defining the electron confining potentials, while keeping the 
tunneling barriers low, thereby depleting the chain, and then doping one or 
more preselected QDs with single electrons by lowering the confining 
potentials and carefully manipulating the tunneling barriers between 
the dots and the external electron reservoir \cite{EHGetc,VHWvBetc}. 
Finally, the system can be isolated by closing the tunnelings between the two 
ends of the chain and the reservoir. To start the evolution, one then 
tunes the interdot tunneling rates to the preselected values. This process 
should, on the one hand, be fast enough on the time scale of $t_{ij}^{-1}$ 
so that no appreciable change in the initial state of the system occurs 
during the switching time $\tau_{\rm sw}$, and, on the other hand, 
adiabatic so that no nonresonant coupling between the dots is induced:
$\ve^{-1}, U^{-1} < \tau_{\rm sw} < t_{ij}^{-1}$.

To examine the influence of disorder due to structure imperfections, gate 
voltage fluctuations \cite{ElTrQD,ReMa,WDFEFTK,spdch}, electron-phonon 
interactions \cite{phonons}, etc., we allow for random fluctuations of the 
intradot energy levels and interdot couplings, which would result in a 
decoherence of the electron wavepacket propagation. Specifically, we model 
the energy levels $\ve_{j}$ and the interdot couplings $t_{ij}$ as Gaussian 
random variables, with mean values $\ve_{0}$, $t_0$ and variances 
$\de \ve,\de t < t_0$, respectively. To account also for a possible 
measurement process, we assume the last dot of the chain to be dissipatively 
coupled to an external single electron detector (SED) with a rate 
$\gamma_{\alpha}=\gamma$ (Fig. \ref{scheme}). Such detection can be 
realized by coupling the last dot to another larger measuring dot at 
lower potential, which is continuously monitored by a nearby quantum point 
contact conductor, realizing a broad-bandwidth single electron transistor 
\cite{ElTrQD,set1,set2}. Then, the electron tunnels out of the chain into 
an effective continuum represented by many available, broad (due to the 
continuous monitoring), empty levels of the measuring dot.

The treatment of this type of problems is amenable to the density operator 
approach \cite{GuPr,WeNa}. In the case of only one 
excess electron present in the system, one has to solve $N^2$ coupled 
differential equations (equivalent to the optical Bloch equations) 
for the density matrix elements. When, however, the initial number 
$n_e$ of excess electrons in the system is larger than one, $N \gg n_e >1$, 
the number of coupled density matrix equations to be solved scales 
approximately as $N^{2n_e}$, which is computationally demanding. 
Having in mind the subsequent application of the method to a chain 
of QDs doped initially with more than one electron (see below), we adopt 
here an alternative but equivalent approach based on the Monte Carlo 
stochastic wavefunctions, which has been well developed and widely 
employed in the context of quantum optics \cite{DCM,DZR,RevQJumps}. 
Thus, for a set of Gaussian random numbers $\ve_{j}$, $t_{ij}$ we propagate 
the wavefunction of the system $\ket{\Psi(\tau)}$ using the effective 
non-hermitian Hamiltonian $H_{\rm eff} = H - \frac{i}{2}\gamma n_N$, 
where $H$ is given by Eq. (\ref{Ham}). This involves solving $\sim N^{n_e}$ 
amplitude equations. The fact that $H_{\rm eff}$ is non-hermitian implies 
that the norm of the wavefunction is not conserved. The propagation is 
interrupted by quantum jumps (corresponding to detector clicks in a real 
experiment) that occur when $||\Psi(\tau)||^2=r$, where $r$ is a uniformly 
distributed random number, between $0$ and $1$. Thus, a quantum jump 
corresponds to the detection of one electron by the detector, while the
loss of the electron from the system projects the initial subspace of 
$N$ QDs sharing $n_e$ electrons, to a subspace of $N$ QDs sharing $n_e-1$ 
electrons. The post-jump wave function of the system is thus given by 
\be
\ket{\Psi}\rightarrow\sum_\alpha
\frac{a_{N\alpha}\ket{\Psi}}{\sqrt{\bra{\Psi}\ak_{N\alpha} 
a_{N\alpha}\ket{\Psi}}},
\label{pjwf}
\ee 
where the denominator ensures its renormalization. A new random number $r$ 
is then generated which determines the subsequent jump event in our 
simulations. The propagation is terminated once the system has reached the 
zero electron state $\ket{0} \equiv \ket{0_1,...0_N}$. This procedure is 
repeated and the results are averaged over a large number of independent 
realizations (trajectories). Thus this method allows one to model both
a single experiment outcome, corresponding to a single trajectory, as well
as the ensemble average measurement, corresponding to the averaging
over many independent trajectories, while the density matrix approach
is capable to adequately describe only the latter. Since in the quantum
computation schemes, one usually deals with a single measurement on the 
qubit \cite{QCI} (with the notable exception of NMR-based QCs), the 
Monte Carlo approach is more suitable (natural) for modeling the readout 
from a QC.

\section{One excess electron in the chain}
\label{sec:1e}

Let us consider first the simplest case of a chain of QDs initially doped 
with one excess electron. Since the Hamiltonian (\ref{Ham}) preserves the 
number of electrons and their spin states, the system is restricted throughout
its evolution to the one-electron Hilbert space ${\cal H}_1$, and more 
precisely to its subspace ${\cal H}_1^\alpha$ pertaining to the initial 
spin state of the excess electron. Thus, the total wavefunction reads 
\be
\ket{\psi_1(\tau)} = \sum_{j,\alpha}^N A_{j}^{\alpha}(\tau)\ket{j_\alpha},
\label{wf1}
\ee
where $\ket{j_\alpha} \equiv \ak_{j\alpha} \ket{0_1,...,0_N}$ denotes the 
state where one electron with spin $\alpha$ is located at the $j$th dot. 
The time-evolution of $\ket{\psi_1(\tau)}$ is governed by the Schr\"odinger 
equation, which yields 
\be
i \frac{d A_j^{\alpha}}{d\tau} = \ve_{j} A_j^{\alpha} 
+ t_{j-1j} A_{j-1}^{\alpha} + t_{jj+1} A_{j+1}^{\alpha} , 
\label{em1e}
\ee 
where $t_{01} = t_{NN+1} =0$ and $\hbar =1$. 
Clearly, the two sets of these amplitude equations, pertaining to the spin up 
and spin down states, are completely equivalent and decoupled from each other,
i.e., the subspaces ${\cal H}_1^\us$ and ${\cal H}_1^\ds$ are closed. 
As a result, if the excess electron is initially prepared in an arbitrary 
superposition of spin up and spin down states, 
$\ket{\psi_1(0)} = A_{j}^{\us}\ket{j_\us}+A_{j}^{\ds}\ket{j_\ds}$, the two 
parts of the wavefunction will evolve symmetrically and independently of 
each other. This assertion is valid as long as one suppresses 
all uncontrollable spin-flip processes due to, e.g., presence of stray 
magnetic fields or spin-phonon coupling, which would otherwise destroy 
any pure spin state of the electron. Note that if the electrons in the 
array are to be used as qubits or carriers of quantum information, preserving 
their spin-coherence over long times becomes vital. 
This may potentially be realized by applying a strong transverse magnetic 
field, that would remove the spin degeneracy and make the spin-flip process 
energetically unfavorable. Experimental measurement of spin-relaxation 
times indicate sub-MHz rates \cite{spdch}.

\begin{figure}[t]
\centerline{\includegraphics[width=8.5cm]{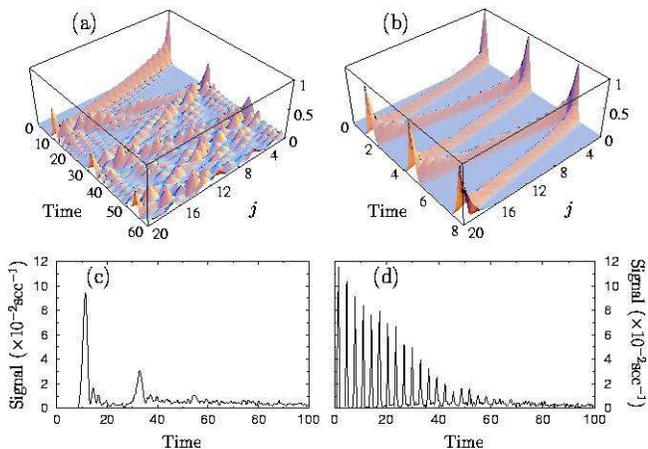}}
\caption{(a), (b) Single-electron transport in a chain of $N=20$ QDs 
with no disorder and dissipation ($\de \ve = \de t = \ga = 0$) for
the case of (a) equal couplings, $t_{jj+1}=t_0$, and (b)
optimal couplings  $t_{jj+1}=t_0 \sqrt{(N-j)j}$, between the dots .
(c),(d) Monte-Carlo simulations of detector response averaged over 5000 
trajectories for the chain with disorder and dissipation 
($\de \ve = 0.1$, $\de t = 0.05$ and $\ga = 0.2$) for the case 
of (a) equal couplings, and (b) optimal couplings between the dots.
All parameters are normalized by $t_0$ and the time is in units of $t_0^{-1}$.
\label{ch_1e}}
\end{figure}
 
In Fig. \ref{ch_1e}(a) we plot the time-evolution of the occupation 
probabilities of $N=20$ QDs along a uniform ($\ve_j = \ve_0, 
t_{ij} = t_0$), isolated from the environment chain, for the initial 
state $\ket{\psi_1(0)}=\ket{1_\alpha}$. The initially well localized 
electron wavepacket seen to propagate along the chain and spread to a 
larger number of dots, while simultaneously splitting into several smaller 
components, which is due to the asymmetry of the initial conditions with
respect to the electron forward and backward propagation. When the leading
edge of the wavepacket is reflected from the end of the chain, its forward 
and backward propagating components begin to interfere with each other and
after several reflections from the boundaries of the chain, this 
interference causes the complete delocalization of the wavepacket 
over the entire chain. 

The results shown in Figs. \ref{ch_1e}(a),(b) have been obtained through the 
numerical solution of Eqs. (\ref{em1e}). These solutions, however, can
be expressed in an analytical form obtainable through the Laplace 
transform. If one eliminates the first term via the transformation 
$A_j^{\alpha}\to A_j^{\alpha} e^{i \ve_j \tau}$, which is equivalent
to the interaction picture, taking the Laplace transform of the set of
Eqs. (\ref{em1e}), leads to 
\[
i s K_j^{\alpha}(\la) - i A_j^{\alpha}(0) = t_0   K_{j-1}^{\alpha}(\la) + 
t_0   K_{j+1}^{\alpha}(\la) ,
\]
where we have assumed equal energies $\ve_0$ and tunneling rates $t_0$.
Considering the case of only dot 1 occupied initially, which means
$A_j^{\alpha}(0)=\de_{1j}$, the quantities $K_j^{\alpha}$ are expressed in
terms of the determinants of the coefficients. For $N$ dots, the determinant
${\cal D}_N(\la)$ is readily shown to obey the recursion relation 
${\cal D}_N(\la)= \la {\cal D}_{N-1}(\la) +t_0^2 {\cal D}_{N-2}(\la)$
with ${\cal D}_0=1$ and ${\cal D}_1=\la$, which identifies it with the
Chebyshev polynomial of the second kind 
${\cal D}_N(\la) = \Pi_{k=1}^{N} (\la-\la_k)$ known to have the roots 
$\la_k = 2 i t_0 \cos \left( \frac{k \pi}{N+1}\right)$. The determinants in 
the numerators of the expressions for $K_j^{\alpha}(\la)$ have similar 
properties. Taking advantage of this connection with the Chebyshev 
polynomials, one can, after some algebra, write the solutions for 
the amplitudes as 
\begin{eqnarray*}
A_j^{\alpha} &=&  \frac{2}{N+2} \sum_{k=1}^{N} 
\exp \left[-i 2 t_0 \tau \cos \left( \frac{k \pi}{N+1}\right)\right] \times \\
& & \times \sin\left( \frac{j k \pi}{N+1}\right) 
\sin\left( \frac{k \pi}{N+1}\right).
\end{eqnarray*}
It is thus evident that the eigenstates of the coupled system oscillate 
with incommensurate frequencies corresponding to the roots $\la_k$ of 
${\cal D}_N$, which in fact become increasingly densely spaced with increasing
$N$. As a consequence, the system never revives fully to its initial state.
This result has also been obtained in an entirely different context of 
driven multilevel systems  \cite{shore1,shore2,shore3}. 

Clearly, it is highly desirable to tailor the parameters of the system
so as to achieve a non-dispersive transfer of the single-electron wavepacket
between the two ends of the chain. This can be accomplished through the
judicious choice of the interdot tunneling matrix elements, namely by choosing
$t_{jj+1}$ according to $t_{jj+1}=t_0 \sqrt{(N-j)j}$, $j=1,...,N-1$
[Fig. \ref{ch_1e}(b)], to be referred to hereafter as optimal coupling. 
Then again, by exploring the properties of the Jacobi polynomials, one 
can obtain an analytic expression for the amplitudes, which in fact is 
even simpler, given by the binomial form
\[
A_j^{\alpha} = \left(\begin{array}{c}
N-1 \\ j-1 \end{array} \right)^{1/2} 
[-i \sin{(t_0 \tau)}]^{(j-1)} \cos{(t_0 \tau)}^{(N-j)} ,
\]
while the eigenstates of the system have commensurate frequencies
$\la_k = t_0 (2k-N-1)$, and the electron wavepacket oscillates in a 
perfectly periodic way between the first and the last dots, whose occupation 
probabilities are given, respectively, by 
$|A_1^{\alpha}|^2 = \cos{(t_0 \tau)}^{2(N-1)}$ and 
$|A_N^{\alpha}|^2 = \sin{(t_0 \tau)}^{2(N-1)}$; a behavior reminiscent of
a two-state system, in a generalized sense.

Having established the conditions for the nondispersive transfer of the 
one-electron wavepacket between the ends of the chain, the next issue is 
the time-scale of the persistence of the coherence in the presence of
disorder and/or fluctuations. In an actual experiment, these effects will
also be reflected in the signal detection, i.e., the measurement by a 
detector connected to one end of the chain. As already mentioned in the
previous Section, this process is conveniently modeled through the
quantum Monte Carlo technique. Although not necessary for the main issue
here, we include it for the sake of completeness, as it represents a
mechanism of dissipation inevitably present in an experiment. Continuing
thus with the case of the chain initially doped with one excess electron, 
$\ket{\psi_1(0)}=\ket{1_\alpha}$, we plot in Figs. \ref{ch_1e}(c),(d) 
the results of Monte-Carlo simulations for the detector signal averaged 
over 5000 trajectories. We chose the amplitudes of the energy-level and 
tunneling-rate fluctuations to be $\de \ve = 0.1 t_0$ and $\de t = 0.05 t_0$, 
respectively, which, for typical experimental parameters \cite{cmnt} 
correspond to $\de \ve \sim 5$ $\mu$eV and $\de t \sim 2.5$ $\mu$eV. 
The decay rate $\ga$ of the electron from the last dot has been taken
to be $\ga = 0.2 t_0 \sim 2.4$ GHz. In Fig. \ref{ch_1e}(c) we show the 
average detector signal for the case of equal mean couplings between the 
dots, while in Fig. \ref{ch_1e}(d) the case of optimal mean couplings is 
illustrated. Comparing these figures with those corresponding to the 
isolated uniform chain [Figs. \ref{ch_1e}(a),(b)], as expected, the 
detector signal is found peaked around times when the electron occupies 
the last dot. Over the time scale of $\tau \sim \de \ve^{-1},\de t^{-1}$ 
the system decoheres significantly due to the inhomogeneity introduced by 
the energy and coupling fluctuations, while the total occupation probability 
of the chain decays roughly according to 
$\sum_j |A_j^{\alpha}|^2 \simeq \exp{(-\ga \tau /N)}$;
the latter being the signature of dissipation.

\section{Two excess electrons in the chain}
\label{sec:2e}

We consider next a chain of QDs that is initially doped with two excess 
electrons. In analogy with the one-electron problem, the evolution of the 
system is restricted to the two-electron sector of the Hilbert space 
${\cal H}_2^{\alpha \beta}$. We assume that the intradot Coulomb repulsion 
is so large, $U \gg \ve_j, t_{ij}$, that the two electrons are forced to 
occupy different dots and thus the states $\ket{j_{\alpha},j_{\beta}}$
corresponding to two electrons at the same dot $j$ are practically forbidden. 
Then the state-vector of the system at an arbitrary time $\tau$ is given by  
\be
\ket{\psi_2(\tau)} = \sum_{\alpha\beta}\sum_{i< j }^N 
B_{ij}^{\alpha \beta}(\tau) \ket{i_\alpha,j_\beta},
\label{wf2}
\ee
where $\ket{i_\alpha,j_\alpha} \equiv 
\ak_{i\alpha} \ak_{j\beta} \ket{0_1,...,0_N}$ denotes the state where two
electrons with spins $\alpha$ and $\beta$ ($\alpha, \beta \in \{\us,\ds\}$) 
are located at the $i$th and the $j$th dots, respectively. The corresponding 
amplitudes obey the following equations of motion
\bea
i \frac{d B_{ij}^{\alpha \beta}}{d \tau} & = & 
(\ve_i + \ve_j + V_{ij}) B_{ij}^{\alpha \beta} 
+ t_{i-1i} B_{i-1j}^{\alpha\beta} + t_{ii+1} B_{i+1j}^{\alpha\beta}
\nonumber \\ & &
+ t_{j-1j} B_{ij-1}^{\alpha\beta} + t_{jj+1} B_{ij+1}^{\alpha\beta} ,
\label{em2e}
\eea
and obviously, the four subspaces (${\cal H}_2^{\us \us}$, 
${\cal H}_2^{\us \ds}$, ${\cal H}_2^{\ds \us}$ and ${\cal H}_2^{\ds \ds}$) of 
${\cal H}_2^{\alpha \beta}$ are governed by equivalent sets of equations.
These sets are decoupled from each other, since the Hamiltonian (\ref{Ham})
does not contain terms that directly couple different spin states. 
Note that since the states $\ket{j_{\alpha},j_{\beta}}$ are excluded
from state-vector (\ref{wf2}), the transient Heisenberg coupling 
$J_0 = 4 t_0^2 /U$ between the spins is not accounted for in our model. 
This is justified provided during the time $\tau$ over which we consider the 
system dynamics the probability of this second-order spin-exchange process 
remains small, $J_0 \tau \ll 1$, which in turn requires that 
$U \gg 4 t_0^2 \tau$ \cite{heis}. Then, due to the linearity of the 
Schr\"odinger equation, throughout the evolution each electron will 
preserve its spin state and the two electrons will not penetrate through 
one another, remaining thus in principle distinguishable. 

\begin{figure}[t]
\centerline{\includegraphics[width=8.5cm]{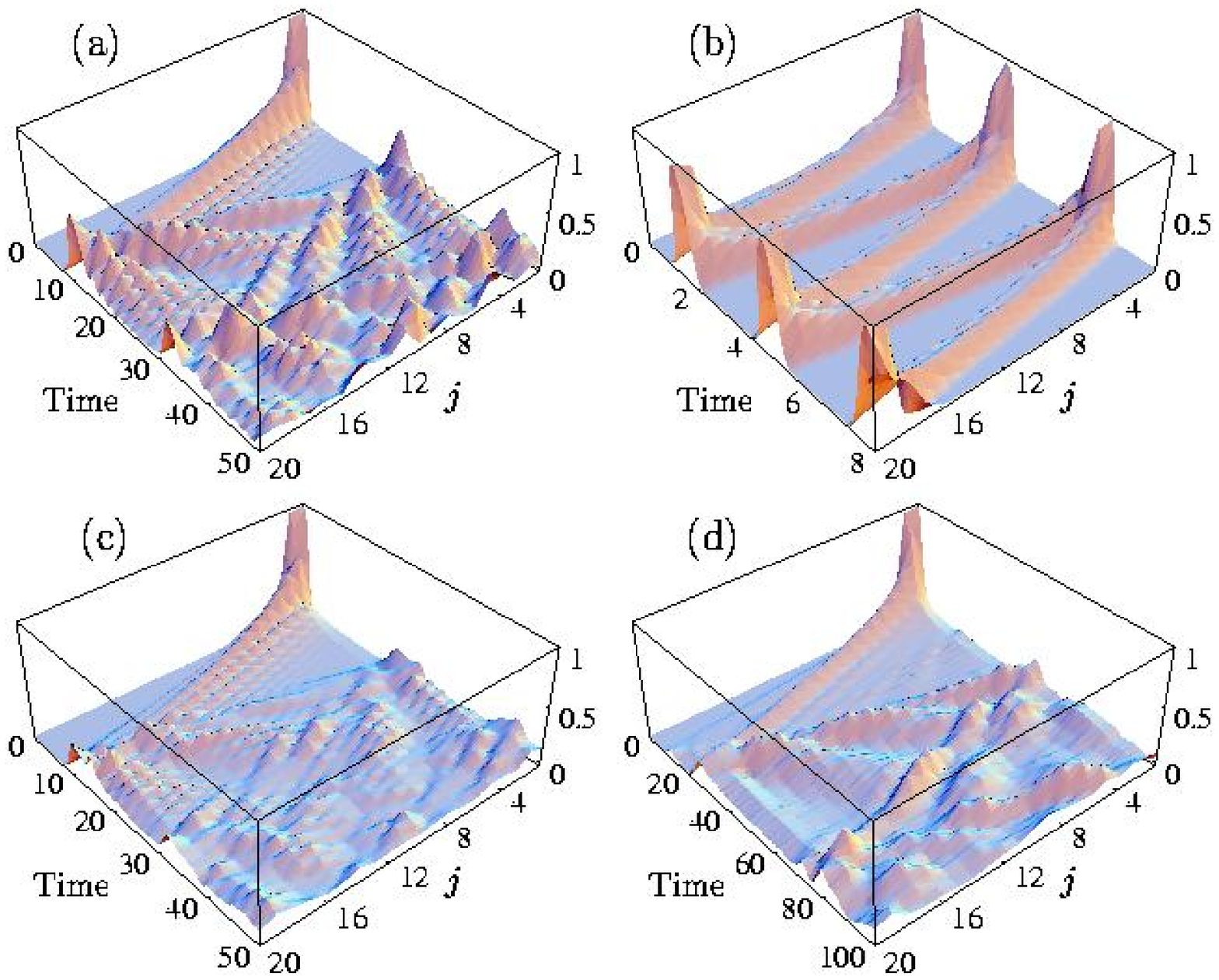}}
\caption{Two-electron transport in a uniform ($\de \ve = \de t = 0$), 
isolated ($\ga = 0$) chain of $N=20$ QDs for the initial state
$\ket{\psi_2(0)}=\ket{1_\alpha,2_\beta}$. 
(a) Equal couplings between the dots, $t_{jj+1}=t_0$, and (b) optimal 
couplings  $t_{jj+1}=t_0 \sqrt{(N-j)j}$ ($j=1,...,N-1$) and no interdot
repulsion, $V=0$. (c) Equal couplings and weak repulsion $V=0.75$, and 
(d) equal couplings and strong repulsion $V=2.5$. Note the different 
scales of the time axis.
\label{2ess}}
\end{figure}

Unlike the one-electron case, useful analytic solutions do not seem easy
to come by in the presence of a second electron, except for special cases. 
Also the analogy with the driven multilevel systems exists no longer.
Obtained through the numerical solutions of Eqs. (\ref{em2e}), in 
Fig. \ref{2ess} we plot the time-evolution of the occupation probabilities 
of the QDs along a uniform, isolated chain for the initial state 
$\ket{\psi_2(0)}=\ket{1_\alpha,2_{\beta}}$. Consider first the case of
vanishing interdot Coulomb repulsion $V = 0$. Similarly to the 
single-electron case, the two-electron wavepacket propagates along the 
chain, spreads to a larger number of dots and, after several reflections
from the boundaries, becomes delocalized over the entire chain, due to
the multiple interference of its forward and backward propagating components
[Fig.\ref{2ess}(a)]. These interference effects can be compensated for by
employing the optimal coupling between the dots. Then every time the 
wavepacket reaches the end of the chain, the two electrons become fully 
localized at the last two dots [Fig.\ref{2ess}(b)]. The system is then
formally (mathematically) equivalent to a chain of $2N-3$ QDs doped with 
one electron. The solution for the amplitudes has the following analytic form
\begin{eqnarray*}
B_{ij}^{\alpha\beta} &=& \left[\frac{(j-i)^2 (N-1)!(N-2)!}
{(i-1)!(j-1)!(N-i)!(N-j)!}\right]^{1/2} \\ & &
[-i \sin{(t_0 \tau)}]^{i+j-3} \cos{(t_0 \tau)}^{2(N-2)-(i+j-3))} .
\end{eqnarray*}
while the system has $2N-3$ distinct, commensurate eigenstates 
$\la_k = t_0 (2k-2N+2)$, where $1 \leq k \leq 2N-3$.

\begin{figure}[t]
\centerline{\includegraphics[width=8.5cm]{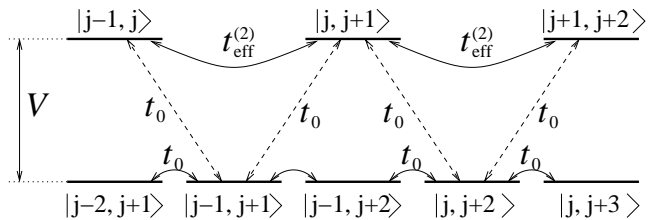}}
\caption{Energy-level diagram for two-electron states in the chain of QDs. 
\label{levels}}
\end{figure}

Consider now the case of a finite near-neighbor interdot repulsion 
$V > 0$ and equal tunneling rates between the dots. The magnitude 
of this repulsion determines the energy mismatch between the states 
$\ket{i_{\alpha},j_{\beta}}$ corresponding to two electrons occupying 
adjacent dots, $j-i=1$, and those corresponding to two electrons 
separated by one or more empty dots, $j-i > 1$. As we show in 
Fig. \ref{2ess}(c), for relatively small interdot repulsion $V=0.75 t_0$, the 
spreading and delocalization of the two-electron wavepacket is accelerated,
which is due to inhomogeneity of the eigenenergies of states 
involved in the system's evolution. For larger interdot repulsion $V=2.5 t_0$,
however, we find a rather counterintuitive behavior of the system: the 
propagation dynamics of the two-electron wavepacket and its delocalization
is slowed down [Fig. \ref{2ess}(d)]. This stems from the fact that when the 
two electrons are localized at adjacent dots, the corresponding states 
$\ket{j_\alpha,(j+1)_\beta}$ are coupled directly only to the states 
$\ket{(j-1)_\alpha,(j+1)_\beta}$ and $\ket{j_\alpha,(j+2)_\beta}$ 
[Fig. \ref{levels}]. But if the interdot repulsion is larger than the 
tunneling rate, $V > t_0$, these transitions are nonresonant and the two 
electrons appear to be {\em bonded} with each other. Then, during the system 
evolution, the latter states are only virtually populated, while the 
second-order process responsible for the direct transition 
$\ket{j_\alpha,(j+1)_\beta} \to \ket{(j+1)_\alpha,(j+2)_\beta}$ is resonant. 
The effective coupling strength of this transition is given by 
$t_{\rm eff}^{(2)} = t_0^2/V < t_0$, and therefore the two-electron wavepacket
propagation is slowed down. Since the energy of the two-electron bonded state 
is larger than that of two separate electrons, the bonded state is not stable 
and non-adiabatic perturbations, such as collisions with the chain boundaries,
gradually destroy it [Fig. \ref{2ess}(d)].

\begin{figure}[t]
\centerline{\includegraphics[width=8.5cm]{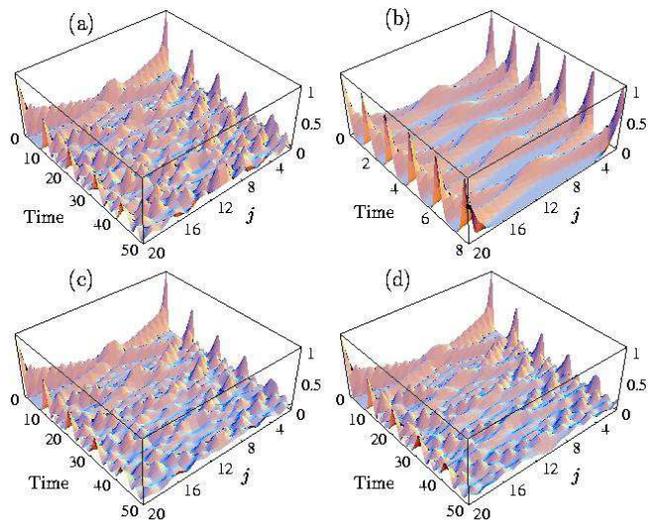}}
\caption{Two-electron transport in a uniform, isolated  chain of $N=20$ 
QDs for the initial state $\ket{\psi_2(0)}=\ket{1_\alpha,20_\beta}$. 
(a) Equal couplings between the dots, $t_{jj+1}=t_0$, and (b) optimal 
couplings  $t_{jj+1}=t_0 \sqrt{(N-j)j}$ ($j=1,...,N-1$) and no interdot
repulsion, $V=0$. (c) Equal couplings and weak repulsion $V=0.75$, and 
(d) equal couplings and strong repulsion $V=2.5$.
\label{2eos}}
\end{figure}

Consider next the time-evolution of the system for the case when the two 
electrons are initially localized at the opposite ends of the chain,
$\ket{\psi_2(0)}=\ket{1_\alpha,N_{\beta}}$ (Fig. \ref{2eos}). Upon 
propagation, the electrons collide with each other in the middle of
the chain and then separate again. For each electron, the presence of the 
other represents an effective time-dependent potential barrier, which 
restricts the number of available dots per electron. Therefore, while after
each collision the system partially revives to its initial state, the overall
spreading and delocalization of the two wavepackets is slightly enhanced
as compared to the case of Fig. \ref{2ess}. Again, choosing the interdot
tunneling rates according to the optimal rates $t_{jj+1}=t_0 \sqrt{(N-j)j}$, 
one can compensate for the interference effects and achieve a perfect 
periodicity in the system dynamics [Fig. \ref{2eos}(b)]. Finally, we note 
that increasing the interdot Coulomb repulsion strength results in 
acceleration of delocalization even further and inhibition of occupation 
of the central dots of the chain where the two electrons collide 
[Fig. \ref{2eos}(c),(d)].

\begin{figure}[t]
\centerline{\includegraphics[width=8.5cm]{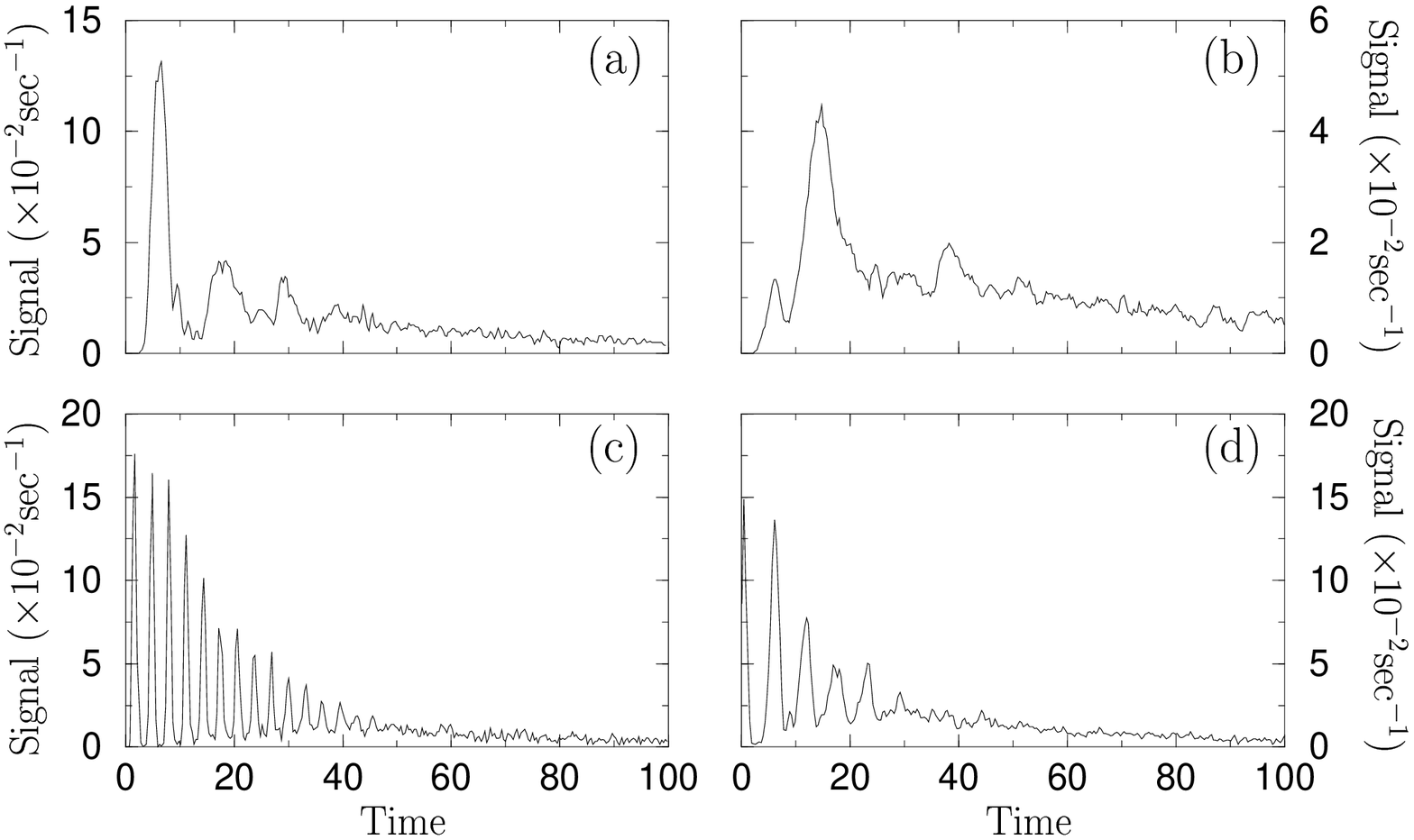}}
\caption{Monte-Carlo simulations of detector response for the 
two-electron transport in a chain of $N=10$ QDs with disorder 
and dissipation ($\de \ve = 0.1$, $\de t = 0.05$ and $\ga = 0.2$), 
averaged over 5000 trajectories. 
In (a), (b) and (c) the initial state is 
$\ket{\psi_2(0)}=\ket{1_\alpha,2_\beta}$, and in (d) the initial state is 
$\ket{\psi_2(0)}=\ket{1_\alpha,10_\beta}$. 
Other parameters are: (a), (d) equal mean couplings between the dots and 
no interdot repulsion, $V=0$; 
(b) equal couplings and strong repulsion, $V=2.5$;  
(c) optimal couplings and no repulsion, $V=0$.
\label{2eqj}}
\end{figure}

We proceed now to the application of the quantum jumps method to the 
two-electron problem. Before the first jump occurs, the evolution of the 
system is restricted to the two-electron Hilbert space 
${\cal H}_2^{\alpha\beta}$. Immediately after the jump, the system is 
projected onto the single-electron Hilbert space ${\cal H}_1^{\alpha}$ 
where it evolves until the second jump event. The total wavefunction 
of the system is thus given by $\ket{\Psi}=\ket{\psi_2}+\ket{\psi_1}+\ket{0}$,
where $\ket{\psi_{1,2}}$ are defined in Eqs. (\ref{wf1}) and (\ref{wf2}), 
respectively. In Fig. \ref{2eqj} we plot the results of our Monte Carlo 
simulations for the detector signal, averaged over 5000 trajectories. 
Due to limitations on the computer CPU time, we simulate a chain of 
$N=10$ QDs, in which the energy and tunneling-rate fluctuations 
and the decay rate from the last dot are taken to be the same as in 
Fig. \ref{ch_1e}(c),(d), $\de \ve = 0.1 t_0$, $\de t = 0.05 t_0$ and 
$\ga = 0.2 t_0$. (We do realize that $N=10-20$ may be demanding for present 
day technology, which, however, continues evolving.) In the two-electron case,
the dynamics of the system is more involved as compared to the single-electron
case, and the interpretation of the behavior of the detector signal is more 
subtle. At small times, while the probability that the first jump has occurred
is not yet very large $||\psi_2(\tau)||^2 \lesssim 1$, one can visualize the 
behavior of the detector signal by comparing the four plots in Fig. \ref{2eqj}
with the corresponding 3D graphs in Figs. \ref{2ess} and \ref{2eos}. One
thus sees that the signal is peaked around times when the population of 
the last dot is maximal. At larger times, however, the averaged over 
many trajectories signal represents an interplay between the two-electron 
propagation before the first jump, and single-electron propagation 
after the jump. Since the jumps occur at random times, the 
single-electron problem at each trajectory has different initial conditions, 
and therefore at larger times the detector signal does not have any 
pronounced structure except for an overall exponential decay with the rate 
$\ga /N$. Thus in Fig. \ref{2eqj}(a) the three pronounced peaks of the 
signal around $\tau \simeq 7, 18$ and $30 \times t_0^{-1}$ correspond to the
arrival of the two-electron wavepacket at the last dot. In Fig. \ref{2eqj}(b),
where we simulate the case of two-electron bonding via strong interdot Coulomb
repulsion $V=2.5 t_0$, the first small peak of the signal around 
$\tau \simeq 7 \times t_0^{-1}$ corresponds to the small unbonded fraction of 
the two-electron wavepacket, while the second larger peak represents the 
delayed arrival of the bonded pair of electrons at the last dot (recall 
that the effective tunneling rate $t_{\rm eff}^{(2)}$ for the bonded 
electron pair is smaller than the single-electron tunneling rate $t_0$). 
Our numerical simulations show that increasing the interdot repulsion does 
indeed strengthen the bonding and reduces the height of the first peak. Next,
in Fig. \ref{2eqj}(c), where we simulate the case of optimal couplings between
the dots, the periodic behavior of the signal at short times reflects the 
periodic arrival of two-electron wavepacket at the last dot. Finally, in
Fig. \ref{2eqj}(d), we illustrate the results for the case when the two 
electrons are initially localized at the opposite ends of the chain. Here, 
before the first jump has occurred, the second electron is effectively 
confined in a chain of length $N/2$ QDs. As a result, the signal initially 
oscillates with twice the frequency as compared to the case of 
Fig. \ref{2eqj}(a), while after the jump, the entire chain becomes 
accessible to the single remaining electron.

\section{Entanglement of two-electrons}
\label{sec:etngl}

Material particles such as electrons represent a promising alternative to 
circumvent communication and detection loopholes that occur during 
the tests of Bell inequalities \cite{MOY}. Despite an ever growing number of 
theoretical proposals \cite{MOY,SarLos,HuSarma_epr,ZXZX}, the experimental 
demonstration of entanglement generation and detection in condensed 
matter systems is still an open issue. One of the hardest steps towards 
such a realization is the necessity of controllable spatial separation 
of the two constituents of an entangled pair \cite{HuSarma_epr}. 
A similar problem would hinder the realization of large-scale solid-state
QCs. One of the main difficulties with the existing proposals for integrated 
QD based QCs \cite{LDV,BLDV,EHGetc,VHWvBetc} is that the qubits 
(electron spins) interact with the nearest neighbors only, and there 
is no efficient way of transferring the information between distant qubits 
(one can {\sc swap} the qubit state with its neighbor qubit, then the latter
with the next neighbor, etc., which is inefficient since each {\sc swap} 
action requires precise dynamical control of the parameters).  
Our intention in this section is to show that use of the optimal 
coupling circumvents these difficulties. 

We adopt here an entanglement scheme (entangler) proposed in \cite{LDV,BLDV},
which is based on the exchange interaction in a double-dot 
system where each QD is occupied by a single electron. When the interdot 
potential barrier is high, the tunneling is negligible, $t_e \simeq 0$, and
the electronic orbitals are well localized at the corresponding dots.
By lowering the interdot tunnel barrier, one can induce a finite overlap 
between the orbitals. Then the two electrons of this ``artificial molecule'' 
will be subject to a transient Heisenberg coupling 
\cite{LDV,BLDV,HuSarma_epr}
\begin{equation}
H_s(\tau) = - J(\tau) \vec{S}_L \cdot \vec{S}_R,
\end{equation}
where $\vec{S}_{L,R}$ are the spin$-1/2$ operators for the left ($L$) and 
right ($R$) QDs, respectively, and $J(\tau) = 4 t_e^2 (\tau)/U$ is the 
effective time-dependent spin exchange constant. If initially the two 
electrons have opposite spin states, $\ket{\phi}=\ket{L_\us,R_\ds}$, 
and the pulsed Heisenberg coupling is applied for a specific duration such 
that $\theta \equiv \int J(\tau) d \tau = \pi/2$, the square root of swap 
action ($\sqrt{\mbox{\sc swap}}$) is realized. Then the resulting state of 
the system is the maximally entangled state 
$\ket{\phi} \to \frac{1}{\sqrt{2}}(\ket{L_\us,R_\ds}+i\ket{L_\ds,R_\us})$.

\begin{figure}
\centerline{\includegraphics[width=7.5cm]{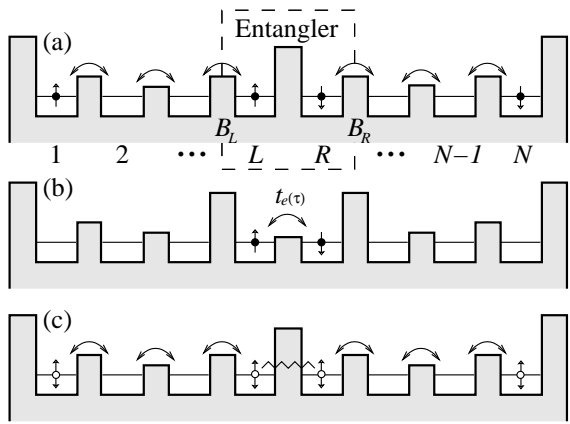}}
\caption{Three step generation of mesoscopically separated entangled electron 
pair. (a) In step 1, employing the optimal coupling for the chains $1-L$ and 
$R-N$, the two electrons are transferred to the entangler. Raising the 
barriers $B_R$ and $B_L$, the entangler is temporarily isolated and the two 
electrons are trapped in dots $L$ and $R$, respectively. 
(b) In step 2 a pulsed Heisenberg exchange coupling is applied, by turning 
on and off the tunneling matrix element $t_e(\tau)$. 
(c) In step 3 the two electrons are transferred back to dots $1$ and $N$.}
\label{EPRfig}
\end{figure}

Thus, a possible scheme, capable of producing mesoscopically separated EPR 
correlated pairs of electrons, relies on the double-dot entangler that 
is embedded in a long chain of $N=2M$ coupled QDs. To be specific, let us 
consider a symmetric structure depicted in Fig. \ref{EPRfig}(a), in which 
the entangler, defined by the barriers $B_L$ and $B_R$, is represented by the 
two middle dots of the chain, $L=M$ and $R=M+1$.  As discussed at the end 
of this Section, the exact location of the entangler is not restrictive 
for the success of the scheme. The entanglement scenario then proceeds in 
three steps illustrated in Figs. \ref{EPRfig}(a)-(c), respectively. During 
the first step [Fig. \ref{EPRfig}(a)], the potential barrier between the dots 
$L$ and $R$ is kept high and the exchange coupling is closed, $t_e=0$. 
Therefore the two parts of the chain, composed of dots 1---$M$ and 
$(M+1)$---$N$, respectively, are decoupled from each other. The two 
unentangled electrons, initially located at dots 1 and $N$, are coherently 
transfered to the entangler by employing the optimal coupling in both parts of
the chain. At time $\tau = \pi/(2 t_0)$, when each electron reaches the 
corresponding dot of the entangler, the barriers $B_L$ and $B_R$ are pulsed 
to higher potentials so as to trap the corresponding electron. Note that, 
since $t_e=0$, the trapping process need not occur simultaneously for both 
electrons. At the second step [Fig. \ref{EPRfig}(b)], the exchange interaction
is applied, by adiabatically turning on and off the tunneling matrix element 
$t_e$ according to 
$t_e(\tau) = t_e^{\rm max} {\rm sech}[(\tau-\tau^{\rm max})/ \Delta\tau]$, 
where $t_e^{\rm max}$ is the peak amplitude of the pulse and $\Delta\tau$ is
its width. These parameters must be chosen such that the pulse area $\theta$
satisfies $\theta = \pi/2$, which yields 
$(t_e^{\rm max})^2 \De \tau = \pi U/16$. At the same time, according to the 
adiabaticity criteria \cite{LDV}, the conditions   
$\De \tau^{-1},t_e^{\rm max} \ll \De \ve,U$ should be satisfied.
Finally, at the third step [Fig. \ref{EPRfig}(c)], the barriers $B_L$ and 
$B_R$ are reset to their initial values and the two electrons propagate along 
the chain and after time $\tau=\pi/(2t_0)$ become fully localized at the 
opposite ends of the chain, at dots $1$ and $N$.

If the two electrons are prepared initially in the opposite spin states, 
$\ket{\psi_2}=\ket{1_\us,N_\ds}$, during the three-step process the 
system will evolve into the maximally entangled state 
$\ket{\psi_2} \to \frac{1}{\sqrt{2}}(\ket{1_\us,N_\ds}+i\ket{1_\ds,N_\us})$.
If, on the other hand, the two electrons are prepared initially in the same 
spin state, $\ket{1_\us,N_\us}$ or $\ket{1_\ds,N_\ds}$, this state will 
remain unchanged (to within the phase factor $e^{i \theta/2}$). 
Finally we note that if, during the second step, the exchange pulse area 
$\theta$ is equal to $\pi$ (instead of $\pi/2$), the two electrons will 
swap their states $\ket{1_\alpha,N_\beta} \to i \ket{1_\beta,N_\alpha}$.

\begin{figure}
\centerline{\includegraphics[width=8.5cm]{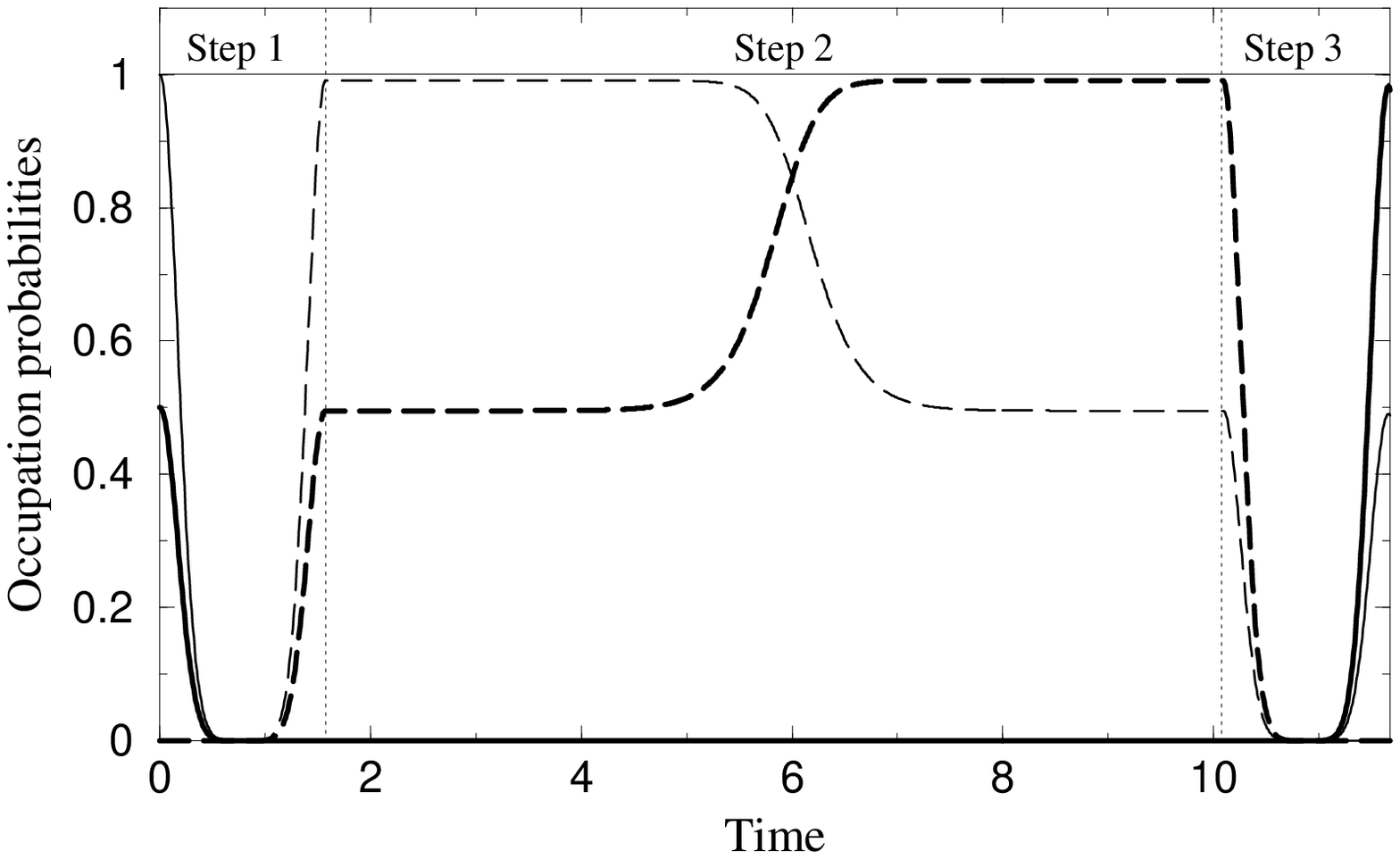}}
\caption{Monte Carlo simulation of entangled state generation in a chain 
of $N=20$ QDs. The initial state is $\ket{\psi_2(0)}=\ket{1_\us,N_\ds}$. 
The occupation probabilities of the two-electron states 
$\ket{1_\us,N_\ds}$ (thin solid line), 
$\ket{L_\us,R_\ds}$ (thin dashed line), 
$\frac{1}{\sqrt{2}}(\ket{L_\us,R_\ds}+i\ket{L_\ds,R_\us})$ 
(thick dashed line), and 
$\frac{1}{\sqrt{2}}(\ket{1_\us,N_\ds}+i\ket{1_\ds,N_\us})$ (thick solid line) 
are plotted as functions of time for the three-step process illustrated in 
Figs. \ref{EPRfig}(a)-(c), respectively. The parameters are 
$t_e^{\rm max} = 6 t_0$, $U= 100 t_0$ ($\De \tau \simeq 0.55 t_0^{-1}$)
and the optimal coupling is employed separately for each subchain 
$1$---$L$ and $R$---$N$. Other parameters are the same as in 
Fig. \ref{2eqj}.}
\label{EPRgraph}
\end{figure}

We have performed Monte Carlo simulations of the above scenario for an even 
total number of nearly identical QDs and the initial state 
$\ket{\psi_2}=\ket{1_\us,N_\ds}$. The evolution of the occupation 
probabilities $|\braket{\phi_i}{\psi_2(\tau)}|^2$ for the two-electron states 
$\ket{\phi_0}=\ket{1_\us,N_\ds}$,
$\ket{\phi_1}=\ket{L_\us,R_\ds}$,
$\ket{\phi_2}=\frac{1}{\sqrt{2}}(\ket{L_\us,R_\ds}+i\ket{L_\ds,R_\us})$, and 
$\ket{\phi_3}=\frac{1}{\sqrt{2}}(\ket{1_\us,N_\ds}+i\ket{1_\ds,N_\us})$ 
during the three stages of the process are plotted in Fig. \ref{EPRgraph}. 
One can see that the fidelity of the entangled state generation at the
end of the process is very high, 
$|\braket{\phi_3}{\psi_2(\tau_{\rm end})}|^2 \simeq 0.98$. 
Employing spin-sensitive single-electron transistors \cite{set1,set2}, 
one could then use this EPR pair to perform a test of 
Bell inequalities with material particles (electrons).  
The controlled two-electron coherent dynamics studied above could be also 
envisioned in an integrated quantum register. Such a register would be 
composed of a large number of sub-registers, each containing two or more 
adjacent qubits represented by spins of single electrons in individual QDs. 
The sub-registers are embedded in a two-dimensional {\it array of empty QDs}.
In combination with the mechanism of controlling the tunnel-coupling
between the dots, this two-dimensional grid would realize a flexible
quantum channel, capable of connecting any pair of qubits within the
register. Thus, to transfer the information, one connects distant 
sub-registers by a chain of empty QDs and arranges the optimal 
tunnel-coupling between the dots to achieve a non-dispersive propagation 
of the qubit, followed by its controlled entanglement or \textsc{swap} 
with a target qubit. Note that this scheme is analogous to a proposal 
for an integrated ion trap based QC \cite{ingrIT}. There, in order to 
circumvent the difficulties associated with single large ion 
trap quantum register, it was proposed to use many small sub-registers, 
each containing only a few ions, and connect these sub-registers to each 
other via controlled qubit (ion) transfer to the interaction region 
(entangler) represented by yet another ion trap. It would seem that
contemplating subregisters consisting of few (as few as two) QDs
is no less practical.

Before closing this section, let us address the issue of location of the 
entangler. Since, as was noted above, the two electrons need not 
arrive to the entangler simultaneously, it is not necessary to position 
the entangler in the middle of the chain or synchronize the propagation 
of the two electrons during the first and the last steps. Actually, the 
entangler can be realized by any pair of dots in the chain, and in particular,
even by the first two dots. Then only the second electron, initially located 
at dot $N$ should be brought into the entangler to interact with the first 
electron.

\section{Conclusions}
\label{sec:concl}
To conclude, we have studied the microscopic dynamics of one or two electron 
transport in a linear array of tunnel-coupled quantum dots. We have shown 
that coherent electron wavepacket propagation in this system reveals a number 
of novel, surprising effects, such as suppression or enhancement of 
interference and spreading, two-electron bonding and collision. We should 
note that the coherent electron wavepacket propagation in arrays of 
tunnel-coupled QDs bears many analogies with spin-wave propagation in 
spin chains \cite{spch}, electromagnetic wave propagation in nonlinear 
periodic media \cite{Kuri2001} and matter-wave propagation in optical 
lattices, where a transition from superfluid to Mott-insulator phase has 
recently been observed \cite{mat_wave1,mat_wave2}. With an unprecedented 
control over system parameters, arrays of QDs allow for studies of numerous 
coherence and correlation effects in many-body physics. As an example, 
dynamical localization of electrons is expected to be observed in the arrays 
with large disorder $\de \ve, \de t \sim t_0$ \cite{SDS,SKDS}.
Our studies are also aimed at possible applications in future quantum 
computation schemes \cite{QCI,RevSteane}, as many solid state quantum 
computer proposals involve arrays of interacting quantum dots 
\cite{LDV,BLDV,EHGetc,VHWvBetc,zanros,LiAra,Tanamoto,BHHFKWSF}. 
We stress that if the electrons in the array are to 
be used as qubits or carriers of quantum information, preserving their 
spin-coherence over long times becomes vital. Two-dimensional arrays of 
electrostatically- and tunnel-interacting quantum dots have also been 
suggested as an efficient tool for analog computation \cite{AnComp}. 
In addition, there is presently growing interest in implementing a quantum 
cellular automata chain \cite{QCA}, in which the computation is performed 
by applying a sequence of unitary transformations to the input state, 
upon its propagation through the chain towards the output. Such a chain 
with tunable couplings between its links would correspond to a universal 
quantum computer, closely resembling the pioneering idea of Feynman 
\cite{Feynman1} to simulate one physical system by another.

\bibliography{qdsref}

\begin{thebibliography}{55}
\expandafter\ifx\csname natexlab\endcsname\relax\def\natexlab#1{#1}\fi
\expandafter\ifx\csname bibnamefont\endcsname\relax
  \def\bibnamefont#1{#1}\fi
\expandafter\ifx\csname bibfnamefont\endcsname\relax
  \def\bibfnamefont#1{#1}\fi
\expandafter\ifx\csname citenamefont\endcsname\relax
  \def\citenamefont#1{#1}\fi
\expandafter\ifx\csname url\endcsname\relax
  \def\url#1{\texttt{#1}}\fi
\expandafter\ifx\csname urlprefix\endcsname\relax\def\urlprefix{URL }\fi
\providecommand{\bibinfo}[2]{#2}
\providecommand{\eprint}[2][]{\url{#2}}

\bibitem[{\citenamefont{Akkermans et~al.}(1995)\citenamefont{Akkermans,
  Montambaux, Pichard, and Zinn-Justin}}]{solst}
\bibinfo{editor}{\bibfnamefont{E.}~\bibnamefont{Akkermans}},
  \bibinfo{editor}{\bibfnamefont{G.}~\bibnamefont{Montambaux}},
  \bibinfo{editor}{\bibfnamefont{J.-L.} \bibnamefont{Pichard}},
  \bibnamefont{and}
  \bibinfo{editor}{\bibfnamefont{J.}~\bibnamefont{Zinn-Justin}}, eds.,
  \emph{\bibinfo{title}{Mesoscopic quantum systems}}
  (\bibinfo{publisher}{North-Holland}, \bibinfo{address}{Amsterdam},
  \bibinfo{year}{1995}).

\bibitem[{\citenamefont{Giannoni et~al.}(1991)\citenamefont{Giannoni, Varos,
  and Zinn-Justin}}]{Qchaos}
\bibinfo{editor}{\bibfnamefont{M.-J.} \bibnamefont{Giannoni}},
  \bibinfo{editor}{\bibfnamefont{A.}~\bibnamefont{Varos}}, \bibnamefont{and}
  \bibinfo{editor}{\bibfnamefont{J.}~\bibnamefont{Zinn-Justin}}, eds.,
  \emph{\bibinfo{title}{Chaos and quantum physics}}
  (\bibinfo{publisher}{North-Holland}, \bibinfo{address}{Amsterdam},
  \bibinfo{year}{1991}).

\bibitem[{\citenamefont{Jaksch et~al.}(1998)\citenamefont{Jaksch, Bruder,
  Cirac, Gardiner, and Zoller}}]{mat_wave1}
\bibinfo{author}{\bibfnamefont{D.}~\bibnamefont{Jaksch}},
  \bibinfo{author}{\bibfnamefont{C.}~\bibnamefont{Bruder}},
  \bibinfo{author}{\bibfnamefont{J.~I.} \bibnamefont{Cirac}},
  \bibinfo{author}{\bibfnamefont{C.~W.} \bibnamefont{Gardiner}},
  \bibnamefont{and} \bibinfo{author}{\bibfnamefont{P.}~\bibnamefont{Zoller}},
  \bibinfo{journal}{Phys. Rev. Lett.} \textbf{\bibinfo{volume}{81}},
  \bibinfo{pages}{3108} (\bibinfo{year}{1998}).

\bibitem[{\citenamefont{Greiner et~al.}(2002)\citenamefont{Greiner, Mandel,
  Esslinger, Haensch, and Bloch}}]{mat_wave2}
\bibinfo{author}{\bibfnamefont{M.}~\bibnamefont{Greiner}},
  \bibinfo{author}{\bibfnamefont{O.}~\bibnamefont{Mandel}},
  \bibinfo{author}{\bibfnamefont{T.}~\bibnamefont{Esslinger}},
  \bibinfo{author}{\bibfnamefont{T.~W.} \bibnamefont{Haensch}},
  \bibnamefont{and} \bibinfo{author}{\bibfnamefont{I.}~\bibnamefont{Bloch}},
  \bibinfo{journal}{Nature} \textbf{\bibinfo{volume}{415}}, \bibinfo{pages}{39}
  (\bibinfo{year}{2002}).

\bibitem[{\citenamefont{Krug and Buchleitner}(2001)}]{andreas}
\bibinfo{author}{\bibfnamefont{A.}~\bibnamefont{Krug}} \bibnamefont{and}
  \bibinfo{author}{\bibfnamefont{A.}~\bibnamefont{Buchleitner}},
  \bibinfo{journal}{Phys. Rev. Lett.} \textbf{\bibinfo{volume}{86}},
  \bibinfo{pages}{3538} (\bibinfo{year}{2001}).

\bibitem[{\citenamefont{Eberly et~al.}(1977)\citenamefont{Eberly, Shore,
  Bialynicka-Birula, and Bialynicki-Birula}}]{shore1}
\bibinfo{author}{\bibfnamefont{J.~H.} \bibnamefont{Eberly}},
  \bibinfo{author}{\bibfnamefont{B.~W.} \bibnamefont{Shore}},
  \bibinfo{author}{\bibfnamefont{Z.}~\bibnamefont{Bialynicka-Birula}},
  \bibnamefont{and}
  \bibinfo{author}{\bibfnamefont{I.}~\bibnamefont{Bialynicki-Birula}},
  \bibinfo{journal}{Phys. Rev. A} \textbf{\bibinfo{volume}{16}},
  \bibinfo{pages}{2038} (\bibinfo{year}{1977}).

\bibitem[{\citenamefont{Bialynicka-Birula
  et~al.}(1977)\citenamefont{Bialynicka-Birula, Bialynicki-Birula, Eberly, and
  Shore}}]{shore2}
\bibinfo{author}{\bibfnamefont{Z.}~\bibnamefont{Bialynicka-Birula}},
  \bibinfo{author}{\bibfnamefont{I.}~\bibnamefont{Bialynicki-Birula}},
  \bibinfo{author}{\bibfnamefont{J.~H.} \bibnamefont{Eberly}},
  \bibnamefont{and} \bibinfo{author}{\bibfnamefont{B.~W.} \bibnamefont{Shore}},
  \bibinfo{journal}{Phys. Rev. A} \textbf{\bibinfo{volume}{16}},
  \bibinfo{pages}{2048} (\bibinfo{year}{1977}).

\bibitem[{\citenamefont{Cook and Shore}(1979)}]{shore3}
\bibinfo{author}{\bibfnamefont{R.}~\bibnamefont{Cook}} \bibnamefont{and}
  \bibinfo{author}{\bibfnamefont{B.~W.} \bibnamefont{Shore}},
  \bibinfo{journal}{Phys. Rev. A} \textbf{\bibinfo{volume}{20}},
  \bibinfo{pages}{539} (\bibinfo{year}{1979}).

\bibitem[{\citenamefont{Kramer and MacKinnon}(1993)}]{anderson}
\bibinfo{author}{\bibfnamefont{B.}~\bibnamefont{Kramer}} \bibnamefont{and}
  \bibinfo{author}{\bibfnamefont{A.}~\bibnamefont{MacKinnon}},
  \bibinfo{journal}{Rep. Prog. Phys.} \textbf{\bibinfo{volume}{56}},
  \bibinfo{pages}{1469} (\bibinfo{year}{1993}).

\bibitem[{\citenamefont{Jacak et~al.}(1998)\citenamefont{Jacak, Hawrylak, and
  Wijs}}]{QDots}
\bibinfo{author}{\bibfnamefont{L.}~\bibnamefont{Jacak}},
  \bibinfo{author}{\bibfnamefont{P.}~\bibnamefont{Hawrylak}}, \bibnamefont{and}
  \bibinfo{author}{\bibfnamefont{A.}~\bibnamefont{Wijs}},
  \emph{\bibinfo{title}{Quantum Dots}} (\bibinfo{publisher}{Springer-Verlag},
  \bibinfo{address}{Berlin}, \bibinfo{year}{1998}).

\bibitem[{\citenamefont{Kouwenhoven et~al.}(1997)\citenamefont{Kouwenhoven,
  Marcus, McEuen, Tarucha, Westervelt, and Wingreen}}]{ElTrQD}
\bibinfo{author}{\bibfnamefont{L.}~\bibnamefont{Kouwenhoven}},
  \bibinfo{author}{\bibfnamefont{C.}~\bibnamefont{Marcus}},
  \bibinfo{author}{\bibfnamefont{P.}~\bibnamefont{McEuen}},
  \bibinfo{author}{\bibfnamefont{S.}~\bibnamefont{Tarucha}},
  \bibinfo{author}{\bibfnamefont{R.}~\bibnamefont{Westervelt}},
  \bibnamefont{and} \bibinfo{author}{\bibfnamefont{N.}~\bibnamefont{Wingreen}},
  in \emph{\bibinfo{booktitle}{Mesoscopic Electron Transport}}, edited by
  \bibinfo{editor}{\bibfnamefont{L.}~\bibnamefont{Sohn}},
  \bibinfo{editor}{\bibfnamefont{L.~P.} \bibnamefont{Kouwenhoven}},
  \bibnamefont{and} \bibinfo{editor}{\bibfnamefont{G.}~\bibnamefont{Sch{\"o}n}}
  (\bibinfo{publisher}{Kluwer Academic}, \bibinfo{address}{Dordrecht},
  \bibinfo{year}{1997}), vol. \bibinfo{volume}{345} of
  \emph{\bibinfo{series}{Series E: Applied Sciences}}, pp.
  \bibinfo{pages}{105--214}.

\bibitem[{\citenamefont{Kouwenhoven et~al.}(2001)\citenamefont{Kouwenhoven,
  Austing, and Tarucha}}]{KAT}
\bibinfo{author}{\bibfnamefont{L.~P.} \bibnamefont{Kouwenhoven}},
  \bibinfo{author}{\bibfnamefont{D.~G.} \bibnamefont{Austing}},
  \bibnamefont{and} \bibinfo{author}{\bibfnamefont{S.}~\bibnamefont{Tarucha}},
  \bibinfo{journal}{Rep. Prog. Phys.} \textbf{\bibinfo{volume}{64}},
  \bibinfo{pages}{701} (\bibinfo{year}{2001}).

\bibitem[{\citenamefont{Reimann and Manninen}(2002)}]{ReMa}
\bibinfo{author}{\bibfnamefont{S.~M.} \bibnamefont{Reimann}} \bibnamefont{and}
  \bibinfo{author}{\bibfnamefont{M.}~\bibnamefont{Manninen}},
  \bibinfo{journal}{Rev. Mod. Phys.} \textbf{\bibinfo{volume}{74}},
  \bibinfo{pages}{1283} (\bibinfo{year}{2002}).

\bibitem[{\citenamefont{van~der Wiel et~al.}(2003)\citenamefont{van~der Wiel,
  Franceschi, Elzerman, Fujisawa, Tarucha, and Kouwenhoven}}]{WDFEFTK}
\bibinfo{author}{\bibfnamefont{W.~G.} \bibnamefont{van~der Wiel}},
  \bibinfo{author}{\bibfnamefont{S.~D.} \bibnamefont{Franceschi}},
  \bibinfo{author}{\bibfnamefont{J.~M.} \bibnamefont{Elzerman}},
  \bibinfo{author}{\bibfnamefont{T.}~\bibnamefont{Fujisawa}},
  \bibinfo{author}{\bibfnamefont{S.}~\bibnamefont{Tarucha}}, \bibnamefont{and}
  \bibinfo{author}{\bibfnamefont{L.~P.} \bibnamefont{Kouwenhoven}},
  \bibinfo{journal}{Rev. Mod. Phys.} \textbf{\bibinfo{volume}{75}},
  \bibinfo{pages}{1} (\bibinfo{year}{2003}).

\bibitem[{\citenamefont{van~der Vaart et~al.}(1995)\citenamefont{van~der Vaart,
  Godijn, Nazarov, Harmans, Mooij, Molenkamp, and Foxon}}]{VGNHMMF}
\bibinfo{author}{\bibfnamefont{N.~C.} \bibnamefont{van~der Vaart}},
  \bibinfo{author}{\bibfnamefont{S.~F.} \bibnamefont{Godijn}},
  \bibinfo{author}{\bibfnamefont{Y.~V.} \bibnamefont{Nazarov}},
  \bibinfo{author}{\bibfnamefont{C.~J. P.~M.} \bibnamefont{Harmans}},
  \bibinfo{author}{\bibfnamefont{J.~E.} \bibnamefont{Mooij}},
  \bibinfo{author}{\bibfnamefont{L.~W.} \bibnamefont{Molenkamp}},
  \bibnamefont{and} \bibinfo{author}{\bibfnamefont{C.~T.} \bibnamefont{Foxon}},
  \bibinfo{journal}{Phys. Rev. Lett.} \textbf{\bibinfo{volume}{74}},
  \bibinfo{pages}{4702} (\bibinfo{year}{1995}).

\bibitem[{\citenamefont{Kouwenhoven et~al.}(1990)\citenamefont{Kouwenhoven,
  Hekking, van Wees, Harmans, Timmering, and Foxon}}]{KHvWHTF}
\bibinfo{author}{\bibfnamefont{L.~P.} \bibnamefont{Kouwenhoven}},
  \bibinfo{author}{\bibfnamefont{F.~W.~J.} \bibnamefont{Hekking}},
  \bibinfo{author}{\bibfnamefont{B.~J.} \bibnamefont{van Wees}},
  \bibinfo{author}{\bibfnamefont{C.~J. P.~M.} \bibnamefont{Harmans}},
  \bibinfo{author}{\bibfnamefont{C.~E.} \bibnamefont{Timmering}},
  \bibnamefont{and} \bibinfo{author}{\bibfnamefont{C.~T.} \bibnamefont{Foxon}},
  \bibinfo{journal}{Phys. Rev. Lett.} \textbf{\bibinfo{volume}{65}},
  \bibinfo{pages}{361} (\bibinfo{year}{1990}).

\bibitem[{\citenamefont{Stafford and {Das Sarma}}(1994)}]{SDS}
\bibinfo{author}{\bibfnamefont{C.~A.} \bibnamefont{Stafford}} \bibnamefont{and}
  \bibinfo{author}{\bibfnamefont{S.}~\bibnamefont{{Das Sarma}}},
  \bibinfo{journal}{Phys. Rev. Lett.} \textbf{\bibinfo{volume}{72}},
  \bibinfo{pages}{3590} (\bibinfo{year}{1994}).

\bibitem[{\citenamefont{Kotlyar et~al.}(1998)\citenamefont{Kotlyar, Stafford,
  and {Das Sarma}}}]{KSDS}
\bibinfo{author}{\bibfnamefont{R.}~\bibnamefont{Kotlyar}},
  \bibinfo{author}{\bibfnamefont{C.~A.} \bibnamefont{Stafford}},
  \bibnamefont{and} \bibinfo{author}{\bibfnamefont{S.}~\bibnamefont{{Das
  Sarma}}}, \bibinfo{journal}{Phys. Rev. B} \textbf{\bibinfo{volume}{58}},
  \bibinfo{pages}{R1746} (\bibinfo{year}{1998}).

\bibitem[{\citenamefont{Stafford et~al.}(1998)\citenamefont{Stafford, Kotlyar,
  and {Das Sarma}}}]{SKDS}
\bibinfo{author}{\bibfnamefont{C.~A.} \bibnamefont{Stafford}},
  \bibinfo{author}{\bibfnamefont{R.}~\bibnamefont{Kotlyar}}, \bibnamefont{and}
  \bibinfo{author}{\bibfnamefont{S.}~\bibnamefont{{Das Sarma}}},
  \bibinfo{journal}{Phys. Rev. B} \textbf{\bibinfo{volume}{58}},
  \bibinfo{pages}{7091} (\bibinfo{year}{1998}).

\bibitem[{\citenamefont{Gurvitz and Prager}(1996)}]{GuPr}
\bibinfo{author}{\bibfnamefont{S.~A.} \bibnamefont{Gurvitz}} \bibnamefont{and}
  \bibinfo{author}{\bibfnamefont{Y.~S.} \bibnamefont{Prager}},
  \bibinfo{journal}{Phys. Rev. B} \textbf{\bibinfo{volume}{53}},
  \bibinfo{pages}{15932} (\bibinfo{year}{1996}).

\bibitem[{\citenamefont{Gurvitz}(1998)}]{Gurvitz}
\bibinfo{author}{\bibfnamefont{S.~A.} \bibnamefont{Gurvitz}},
  \bibinfo{journal}{Phys. Rev. B} \textbf{\bibinfo{volume}{57}},
  \bibinfo{pages}{6602} (\bibinfo{year}{1998}).

\bibitem[{\citenamefont{Wegewijs and Nazarov}(1999)}]{WeNa}
\bibinfo{author}{\bibfnamefont{M.~R.} \bibnamefont{Wegewijs}} \bibnamefont{and}
  \bibinfo{author}{\bibfnamefont{Y.~V.} \bibnamefont{Nazarov}},
  \bibinfo{journal}{Phys. Rev. B} \textbf{\bibinfo{volume}{60}},
  \bibinfo{pages}{14318} (\bibinfo{year}{1999}).

\bibitem[{\citenamefont{Kang et~al.}(1997)\citenamefont{Kang, Cha, and
  Yang}}]{KChY}
\bibinfo{author}{\bibfnamefont{K.}~\bibnamefont{Kang}},
  \bibinfo{author}{\bibfnamefont{M.-C.} \bibnamefont{Cha}}, \bibnamefont{and}
  \bibinfo{author}{\bibfnamefont{S.-R.~E.} \bibnamefont{Yang}},
  \bibinfo{journal}{Phys. Rev. B} \textbf{\bibinfo{volume}{56}},
  \bibinfo{pages}{R4344} (\bibinfo{year}{1997}).

\bibitem[{\citenamefont{Yu et~al.}(1998)\citenamefont{Yu, Johnson, and
  Heinzel}}]{YJH}
\bibinfo{author}{\bibfnamefont{Z.}~\bibnamefont{Yu}},
  \bibinfo{author}{\bibfnamefont{A.~T.} \bibnamefont{Johnson}},
  \bibnamefont{and} \bibinfo{author}{\bibfnamefont{T.}~\bibnamefont{Heinzel}},
  \bibinfo{journal}{Phys. Rev. B} \textbf{\bibinfo{volume}{58}},
  \bibinfo{pages}{13830} (\bibinfo{year}{1998}).

\bibitem[{\citenamefont{Elzerman et~al.}(2003)\citenamefont{Elzerman, Hanson,
  Greidanus, {Willems van Beveren}, {De Franceschi}, Vandersypen, Tarucha, and
  Kouwenhoven}}]{EHGetc}
\bibinfo{author}{\bibfnamefont{J.~M.} \bibnamefont{Elzerman}},
  \bibinfo{author}{\bibfnamefont{R.}~\bibnamefont{Hanson}},
  \bibinfo{author}{\bibfnamefont{J.~S.} \bibnamefont{Greidanus}},
  \bibinfo{author}{\bibfnamefont{L.~H.} \bibnamefont{{Willems van Beveren}}},
  \bibinfo{author}{\bibfnamefont{S.}~\bibnamefont{{De Franceschi}}},
  \bibinfo{author}{\bibfnamefont{L.~M.~K.} \bibnamefont{Vandersypen}},
  \bibinfo{author}{\bibfnamefont{S.}~\bibnamefont{Tarucha}}, \bibnamefont{and}
  \bibinfo{author}{\bibfnamefont{L.~P.} \bibnamefont{Kouwenhoven}},
  \bibinfo{journal}{Phys. Rev. B} \textbf{\bibinfo{volume}{67}},
  \bibinfo{pages}{161308(R)} (\bibinfo{year}{2003}).

\bibitem[{\citenamefont{Vandersypen et~al.}(2002)\citenamefont{Vandersypen,
  Hanson, {Willems van Beveren}, Elzerman, Greidanus, {De Franceschi}, and
  Kouwenhoven}}]{VHWvBetc}
\bibinfo{author}{\bibfnamefont{L.}~\bibnamefont{Vandersypen}},
  \bibinfo{author}{\bibfnamefont{R.}~\bibnamefont{Hanson}},
  \bibinfo{author}{\bibfnamefont{L.}~\bibnamefont{{Willems van Beveren}}},
  \bibinfo{author}{\bibfnamefont{J.}~\bibnamefont{Elzerman}},
  \bibinfo{author}{\bibfnamefont{J.}~\bibnamefont{Greidanus}},
  \bibinfo{author}{\bibfnamefont{S.}~\bibnamefont{{De Franceschi}}},
  \bibnamefont{and}
  \bibinfo{author}{\bibfnamefont{L.}~\bibnamefont{Kouwenhoven}}, in
  \emph{\bibinfo{booktitle}{Quantum Computing and Quantum Bits in Mesoscopic
  Systems}} (\bibinfo{publisher}{Kluwer Academic},
  \bibinfo{address}{Dordrecht}, \bibinfo{year}{2002}).

\bibitem[{\citenamefont{Nielsen and Chuang}(2000)}]{QCI}
\bibinfo{author}{\bibfnamefont{M.~A.} \bibnamefont{Nielsen}} \bibnamefont{and}
  \bibinfo{author}{\bibfnamefont{I.~L.} \bibnamefont{Chuang}},
  \emph{\bibinfo{title}{Quantum Computation and Quantum Information}}
  (\bibinfo{publisher}{Cambridge University Press},
  \bibinfo{address}{Cambridge}, \bibinfo{year}{2000}).

\bibitem[{\citenamefont{Steane}(1998)}]{RevSteane}
\bibinfo{author}{\bibfnamefont{A.}~\bibnamefont{Steane}},
  \bibinfo{journal}{Rep. Prog. Phys.} \textbf{\bibinfo{volume}{61}},
  \bibinfo{pages}{117} (\bibinfo{year}{1998}).

\bibitem[{\citenamefont{Loss and DiVincenzo}(1998)}]{LDV}
\bibinfo{author}{\bibfnamefont{D.}~\bibnamefont{Loss}} \bibnamefont{and}
  \bibinfo{author}{\bibfnamefont{D.~P.} \bibnamefont{DiVincenzo}},
  \bibinfo{journal}{Phys. Rev. A} \textbf{\bibinfo{volume}{57}},
  \bibinfo{pages}{120} (\bibinfo{year}{1998}).

\bibitem[{\citenamefont{Burkard et~al.}(1999)\citenamefont{Burkard, Loss, and
  DiVincenzo}}]{BLDV}
\bibinfo{author}{\bibfnamefont{G.}~\bibnamefont{Burkard}},
  \bibinfo{author}{\bibfnamefont{D.}~\bibnamefont{Loss}}, \bibnamefont{and}
  \bibinfo{author}{\bibfnamefont{D.~P.} \bibnamefont{DiVincenzo}},
  \bibinfo{journal}{Phys. Rev. B} \textbf{\bibinfo{volume}{59}},
  \bibinfo{pages}{2070} (\bibinfo{year}{1999}).

\bibitem[{\citenamefont{Zanardi and Rossi}(1998)}]{zanros}
\bibinfo{author}{\bibfnamefont{P.}~\bibnamefont{Zanardi}} \bibnamefont{and}
  \bibinfo{author}{\bibfnamefont{F.}~\bibnamefont{Rossi}},
  \bibinfo{journal}{Phys. Rev. Lett.} \textbf{\bibinfo{volume}{81}},
  \bibinfo{pages}{4752} (\bibinfo{year}{1998}).

\bibitem[{\citenamefont{Li and Arakawa}(2000)}]{LiAra}
\bibinfo{author}{\bibfnamefont{X.-Q.} \bibnamefont{Li}} \bibnamefont{and}
  \bibinfo{author}{\bibfnamefont{Y.}~\bibnamefont{Arakawa}},
  \bibinfo{journal}{Phys. Rev. A} \textbf{\bibinfo{volume}{63}},
  \bibinfo{pages}{012302} (\bibinfo{year}{2000}).

\bibitem[{\citenamefont{Tanamoto}(2000)}]{Tanamoto}
\bibinfo{author}{\bibfnamefont{T.}~\bibnamefont{Tanamoto}},
  \bibinfo{journal}{Phys. Rev. A} \textbf{\bibinfo{volume}{61}},
  \bibinfo{pages}{022305} (\bibinfo{year}{2000}).

\bibitem[{\citenamefont{Bayer et~al.}(2001)\citenamefont{Bayer, Hawrylak,
  Hinzer, Fafard, Korkusinski, Wasilewski, Stern, and Forchel}}]{BHHFKWSF}
\bibinfo{author}{\bibfnamefont{M.}~\bibnamefont{Bayer}},
  \bibinfo{author}{\bibfnamefont{P.}~\bibnamefont{Hawrylak}},
  \bibinfo{author}{\bibfnamefont{K.}~\bibnamefont{Hinzer}},
  \bibinfo{author}{\bibfnamefont{S.}~\bibnamefont{Fafard}},
  \bibinfo{author}{\bibfnamefont{M.}~\bibnamefont{Korkusinski}},
  \bibinfo{author}{\bibfnamefont{Z.~R.} \bibnamefont{Wasilewski}},
  \bibinfo{author}{\bibfnamefont{O.}~\bibnamefont{Stern}}, \bibnamefont{and}
  \bibinfo{author}{\bibfnamefont{A.}~\bibnamefont{Forchel}},
  \bibinfo{journal}{Science} \textbf{\bibinfo{volume}{291}},
  \bibinfo{pages}{451} (\bibinfo{year}{2001}).

\bibitem[{\citenamefont{{\it et al.}}(1987)}]{ising}
\bibinfo{author}{\bibfnamefont{R.~R.~E.} \bibnamefont{{\it et al.}}},
  \emph{\bibinfo{title}{Principles of nuclear magnetic resonance in one and two
  dimensions}} (\bibinfo{publisher}{Clarendon}, \bibinfo{address}{Oxford},
  \bibinfo{year}{1987}).

\bibitem[{\citenamefont{Takagahara}(1996)}]{phonons}
\bibinfo{author}{\bibfnamefont{T.}~\bibnamefont{Takagahara}},
  \bibinfo{journal}{J. Lumin.} \textbf{\bibinfo{volume}{70}},
  \bibinfo{pages}{129} (\bibinfo{year}{1996}).

\bibitem[{\citenamefont{Dalibard et~al.}(1992)\citenamefont{Dalibard, Castin,
  and M{\o}lmer}}]{DCM}
\bibinfo{author}{\bibfnamefont{J.}~\bibnamefont{Dalibard}},
  \bibinfo{author}{\bibfnamefont{Y.}~\bibnamefont{Castin}}, \bibnamefont{and}
  \bibinfo{author}{\bibfnamefont{K.}~\bibnamefont{M{\o}lmer}},
  \bibinfo{journal}{Phys. Rev. Lett.} \textbf{\bibinfo{volume}{68}},
  \bibinfo{pages}{580} (\bibinfo{year}{1992}).

\bibitem[{\citenamefont{Dum et~al.}(1992)\citenamefont{Dum, Zoller, and
  Ritsch}}]{DZR}
\bibinfo{author}{\bibfnamefont{R.}~\bibnamefont{Dum}},
  \bibinfo{author}{\bibfnamefont{P.}~\bibnamefont{Zoller}}, \bibnamefont{and}
  \bibinfo{author}{\bibfnamefont{H.}~\bibnamefont{Ritsch}},
  \bibinfo{journal}{Phys. Rev. A} \textbf{\bibinfo{volume}{45}},
  \bibinfo{pages}{4879} (\bibinfo{year}{1992}).

\bibitem[{\citenamefont{Plenio and Knight}(1998)}]{RevQJumps}
\bibinfo{author}{\bibfnamefont{M.~B.} \bibnamefont{Plenio}} \bibnamefont{and}
  \bibinfo{author}{\bibfnamefont{P.~L.} \bibnamefont{Knight}},
  \bibinfo{journal}{Rev. Mod. Phys.} \textbf{\bibinfo{volume}{70}},
  \bibinfo{pages}{101} (\bibinfo{year}{1998}).

\bibitem[{\citenamefont{Hu and {Das Sarma}}(2001)}]{HDS}
\bibinfo{author}{\bibfnamefont{X.}~\bibnamefont{Hu}} \bibnamefont{and}
  \bibinfo{author}{\bibfnamefont{S.}~\bibnamefont{{Das Sarma}}},
  \bibinfo{journal}{Phys. Rev. A} \textbf{\bibinfo{volume}{64}},
  \bibinfo{pages}{042312} (\bibinfo{year}{2001}).

\bibitem[{cmn()}]{cmnt}
\bibinfo{note}{Typically, in $\sim 50$ nm size GaAs/AlGaAs QDs, separated from
  each other by $\sim 100$ nm, one has $t_{ij} \sim 0.05$ meV, $\De \ve \sim
  0.4$ meV, $U \sim 10$ meV, and at dilution-refrigerator temperatures $T \sim
  2-10$ mK the thermal energy is $k_{\rm B} T \sim 0.2-1$ $\mu$eV
  \cite{ElTrQD}}.

\bibitem[{\citenamefont{Fujisawa et~al.}(2001)\citenamefont{Fujisawa, Tokura,
  and Hirayama}}]{spdch}
\bibinfo{author}{\bibfnamefont{T.}~\bibnamefont{Fujisawa}},
  \bibinfo{author}{\bibfnamefont{Y.}~\bibnamefont{Tokura}}, \bibnamefont{and}
  \bibinfo{author}{\bibfnamefont{Y.}~\bibnamefont{Hirayama}},
  \bibinfo{journal}{Phys. Rev. B} \textbf{\bibinfo{volume}{63}},
  \bibinfo{pages}{R081304} (\bibinfo{year}{2001}).

\bibitem[{\citenamefont{Schoelkopf et~al.}(1998)\citenamefont{Schoelkopf,
  Wahlgren, Kozhevnikov, Delsing, and Prober}}]{set1}
\bibinfo{author}{\bibfnamefont{R.~J.} \bibnamefont{Schoelkopf}},
  \bibinfo{author}{\bibfnamefont{P.}~\bibnamefont{Wahlgren}},
  \bibinfo{author}{\bibfnamefont{A.~A.} \bibnamefont{Kozhevnikov}},
  \bibinfo{author}{\bibfnamefont{P.}~\bibnamefont{Delsing}}, \bibnamefont{and}
  \bibinfo{author}{\bibfnamefont{D.~E.} \bibnamefont{Prober}},
  \bibinfo{journal}{Science} \textbf{\bibinfo{volume}{280}},
  \bibinfo{pages}{1238} (\bibinfo{year}{1998}).

\bibitem[{\citenamefont{Lu et~al.}(2003)\citenamefont{Lu, Ji, Pfeiffer, West,
  and Rimberg}}]{set2}
\bibinfo{author}{\bibfnamefont{W.}~\bibnamefont{Lu}},
  \bibinfo{author}{\bibfnamefont{Z.}~\bibnamefont{Ji}},
  \bibinfo{author}{\bibfnamefont{L.}~\bibnamefont{Pfeiffer}},
  \bibinfo{author}{\bibfnamefont{K.~W.} \bibnamefont{West}}, \bibnamefont{and}
  \bibinfo{author}{\bibfnamefont{A.~J.} \bibnamefont{Rimberg}},
  \bibinfo{journal}{Nature} \textbf{\bibinfo{volume}{423}},
  \bibinfo{pages}{422} (\bibinfo{year}{2003}).

\bibitem[{hei()}]{heis}
\bibinfo{note}{In the opposite limit of large and controllable Heisenberg-type
  spin-exchange coupling between pairs of electrons at adjacent sites, one can
  realize a controlled entanglement between two qubits, each represented by a
  spin state of the corresponding electron \cite{LDV}}.

\bibitem[{\citenamefont{Maitre et~al.}(2000)\citenamefont{Maitre, Oliver, and
  Yamamoto}}]{MOY}
\bibinfo{author}{\bibfnamefont{X.}~\bibnamefont{Maitre}},
  \bibinfo{author}{\bibfnamefont{W.~D.} \bibnamefont{Oliver}},
  \bibnamefont{and} \bibinfo{author}{\bibfnamefont{Y.}~\bibnamefont{Yamamoto}},
  \bibinfo{journal}{Physica E} \textbf{\bibinfo{volume}{6}},
  \bibinfo{pages}{301} (\bibinfo{year}{2000}).

\bibitem[{\citenamefont{Saraga and Loss}(2003)}]{SarLos}
\bibinfo{author}{\bibfnamefont{D.~S.} \bibnamefont{Saraga}} \bibnamefont{and}
  \bibinfo{author}{\bibfnamefont{D.}~\bibnamefont{Loss}},
  \bibinfo{journal}{Phys. Rev. Lett.} \textbf{\bibinfo{volume}{90}},
  \bibinfo{pages}{166803} (\bibinfo{year}{2003}).

\bibitem[{\citenamefont{Hu and {Das Sarma}}()}]{HuSarma_epr}
\bibinfo{author}{\bibfnamefont{X.}~\bibnamefont{Hu}} \bibnamefont{and}
  \bibinfo{author}{\bibfnamefont{S.}~\bibnamefont{{Das Sarma}}},
  \emph{\bibinfo{title}{Double quantum dot turnstile as an electron spin
  entangler}}, \eprint{cond-mat/0307024}.

\bibitem[{\citenamefont{Zhang et~al.}()\citenamefont{Zhang, Xue, Zhao, and
  Xie}}]{ZXZX}
\bibinfo{author}{\bibfnamefont{P.}~\bibnamefont{Zhang}},
  \bibinfo{author}{\bibfnamefont{Q.-K.} \bibnamefont{Xue}},
  \bibinfo{author}{\bibfnamefont{X.-G.} \bibnamefont{Zhao}}, \bibnamefont{and}
  \bibinfo{author}{\bibfnamefont{X.}~\bibnamefont{Xie}},
  \emph{\bibinfo{title}{Generation of spatially-separated spin entanglement in
  a triple quantum dot system}}, \eprint{cond-mat/0307037}.

\bibitem[{\citenamefont{Kielpinski et~al.}(2002)\citenamefont{Kielpinski,
  Monroe, and Wineland}}]{ingrIT}
\bibinfo{author}{\bibfnamefont{D.}~\bibnamefont{Kielpinski}},
  \bibinfo{author}{\bibfnamefont{C.}~\bibnamefont{Monroe}}, \bibnamefont{and}
  \bibinfo{author}{\bibfnamefont{D.~J.} \bibnamefont{Wineland}},
  \bibinfo{journal}{Nature} \textbf{\bibinfo{volume}{417}},
  \bibinfo{pages}{709} (\bibinfo{year}{2002}).

\bibitem[{\citenamefont{Cavanagh et~al.}(1996)\citenamefont{Cavanagh,
  Fairbrother, Palmer, and Skelton}}]{spch}
\bibinfo{author}{\bibfnamefont{J.}~\bibnamefont{Cavanagh}},
  \bibinfo{author}{\bibfnamefont{W.~J.} \bibnamefont{Fairbrother}},
  \bibinfo{author}{\bibfnamefont{A.~G.} \bibnamefont{Palmer}},
  \bibnamefont{and} \bibinfo{author}{\bibfnamefont{N.~J.}
  \bibnamefont{Skelton}}, \emph{\bibinfo{title}{Protein {NMR} Spectroscopy:
  {P}rinciples and Practice}} (\bibinfo{publisher}{Academic},
  \bibinfo{address}{San Diego}, \bibinfo{year}{1996}).

\bibitem[{\citenamefont{Kurizki et~al.}(2001)\citenamefont{Kurizki, Kozhekin,
  Opatrny, and Malomed}}]{Kuri2001}
\bibinfo{author}{\bibfnamefont{G.}~\bibnamefont{Kurizki}},
  \bibinfo{author}{\bibfnamefont{A.}~\bibnamefont{Kozhekin}},
  \bibinfo{author}{\bibfnamefont{T.}~\bibnamefont{Opatrny}}, \bibnamefont{and}
  \bibinfo{author}{\bibfnamefont{B.}~\bibnamefont{Malomed}}, in
  \emph{\bibinfo{booktitle}{Progress in Optics}}, edited by
  \bibinfo{editor}{\bibfnamefont{E.}~\bibnamefont{Wolf}}
  (\bibinfo{publisher}{Elsevier}, \bibinfo{address}{North-Holland},
  \bibinfo{year}{2001}), vol. \bibinfo{volume}{{XXXXII}}, pp.
  \bibinfo{pages}{93--146}.

\bibitem[{\citenamefont{Wu et~al.}(1999)\citenamefont{Wu, Lee, Amemiya, and
  Yasunaga}}]{AnComp}
\bibinfo{author}{\bibfnamefont{N.-J.} \bibnamefont{Wu}},
  \bibinfo{author}{\bibfnamefont{H.}~\bibnamefont{Lee}},
  \bibinfo{author}{\bibfnamefont{Y.}~\bibnamefont{Amemiya}}, \bibnamefont{and}
  \bibinfo{author}{\bibfnamefont{H.}~\bibnamefont{Yasunaga}},
  \bibinfo{journal}{IEICE Trans. Electron.} \textbf{\bibinfo{volume}{E82C}},
  \bibinfo{pages}{1623} (\bibinfo{year}{1999}).

\bibitem[{\citenamefont{Lent et~al.}(1993)\citenamefont{Lent, Tougaw, and
  Porod}}]{QCA}
\bibinfo{author}{\bibfnamefont{C.~S.} \bibnamefont{Lent}},
  \bibinfo{author}{\bibfnamefont{P.~D.} \bibnamefont{Tougaw}},
  \bibnamefont{and} \bibinfo{author}{\bibfnamefont{W.}~\bibnamefont{Porod}},
  \bibinfo{journal}{Appl. Phys. Lett.} \textbf{\bibinfo{volume}{92}},
  \bibinfo{pages}{714} (\bibinfo{year}{1993}).

\bibitem[{\citenamefont{Feynman}(1982)}]{Feynman1}
\bibinfo{author}{\bibfnamefont{R.~P.} \bibnamefont{Feynman}},
  \bibinfo{journal}{Int. J. Theor. Phys.} \textbf{\bibinfo{volume}{21}},
  \bibinfo{pages}{467} (\bibinfo{year}{1982}).

\end{thebibliography}
\bibliographystyle{apsrev}

\newpage

\end{document}